\documentclass[twocolumn,tighten]{aastex63}

\usepackage{times,natbib,graphicx,amsmath,multirow,xspace}
\usepackage{xcolor}
\usepackage{lineno}
\newcommand{\nustar}{\textit{NuSTAR}\xspace}

\newcommand{\nicer}{\textit{NICER}\xspace}

\newcommand{\msun}{$M_{\odot}$\xspace}

\newcommand{\fluxcgs}{ergs~s$^{-1}$~cm$^{-2}$\xspace}

\newcommand{\rin}{$R_{\rm in}$\xspace}
\newcommand{\rg}{$R_{g}$\xspace}
\newcommand{\risco}{$R_{\mathrm{ISCO}}$\xspace}
\newcommand{\cps}{counts s$^{-1}$\xspace}
\newcommand{\relxill}{{\sc relxill}\xspace}
\newcommand{\relxillns}{{\sc relxillns}\xspace}
\newcommand{\rfxconv}{{\sc rfxconv}\xspace}
\newcommand{\source}{\mbox{Cyg X-2}\xspace}

\shorttitle{NuSTAR and NICER observations of Cyg X-2}
\shortauthors{Ludlam et al.}

\begin{document}

\title{Radius constraints from reflection modeling of Cygnus X-2 with NuSTAR and NICER}

\correspondingauthor{R. M. Ludlam}
\email{rmludlam@caltech.edu}

\author[0000-0002-8961-939X]{R.~M.~Ludlam}\thanks{NASA Einstein Fellow}
\affiliation{Cahill Center for Astronomy and Astrophysics, California Institute of Technology, Pasadena, CA 91125, USA}

\author[0000-0002-8294-9281]{E.~M.~Cackett}
\affiliation{Department of Physics \& Astronomy, Wayne State University, 666 West Hancock Street, Detroit, MI 48201, USA}

\author[0000-0003-3828-2448]{J.~A.~Garc\'{i}a}
\affiliation{Cahill Center for Astronomy and Astrophysics, California Institute of Technology, Pasadena, CA 91125, USA}

\author{J.~M.~Miller}
\affiliation{Department of Astronomy, University of Michigan, 1085 South University Ave, Ann Arbor, MI 48109-1107, USA}

\author[0000-0002-5041-3079]{A.~L.~Stevens}
\affiliation{Department of Astronomy, University of Michigan, 1085 South University Ave, Ann Arbor, MI 48109-1107, USA}
\affiliation{Department of Physics \& Astronomy, Michigan State University, 567 Wilson Road, East Lansing, MI48824, USA}

\author[0000-0002-9378-4072]{A.~C.~Fabian}
\affiliation{Institute of Astronomy, Madingley Road, Cambridge CB3 0HA, UK}

\author[0000-0001-8371-2713]{J.~Homan}
\affiliation{Eureka Scientific, Inc., 2452 Delmer Street, Oakland, CA 94602, USA}

\author[0000-0002-0940-6563]{M.~Ng}
\affiliation{MIT Kavli Institute for Astrophysics and Space Research, Massachusetts Institute of Technology, Cambridge, MA 02139, USA}

\author{S. Guillot}
\affiliation{ CNRS, IRAP, 9 avenue du Colonel Roche, BP 44346, F-31028 Toulouse Cedex 4, France}
\affiliation{Universit\'{e} de Toulouse, CNES, UPS-OMP, F-31028 Toulouse, France}

\author{D.~J.~K.~ Buisson}
\affiliation{Department of Physics and Astronomy, University of Southampton, Highfield, Southampton, SO17 1BJ, UK}

\author{D.~Chakrabarty}
\affiliation{MIT Kavli Institute for Astrophysics and Space Research, Massachusetts Institute of Technology, Cambridge, MA 02139, USA}

\begin{abstract}
We present a spectral analysis of \nustar and \nicer observations of the luminous, persistently accreting neutron star (NS) low-mass X-ray binary Cygnus X-2. 
The data were divided into different branches that the source traces out on the Z-track of the X-ray color-color diagram; namely the horizontal branch, normal branch, and the vertex between the two. 
The X-ray continuum spectrum was modeled in two different ways that produced a comparable quality fit.
The spectra showed clear evidence of a reflection component in the form of a broadened Fe~K line, as well as a lower energy emission feature near 1~keV likely due to an ionized plasma located far from the innermost accretion disk. 
We account for the reflection spectrum with two independent models (\relxillns and {\sc rdblur*rfxconv}). 
The inferred inclination is in agreement with earlier estimates from optical observations of ellipsoidal light curve modeling (\relxillns: $i=67^{\circ}\pm4^{\circ}$, {\sc rdblur*rfxconv}: $i=60^{\circ}\pm10^{\circ}$). 
The inner disk radius remains close to the NS (\rin~$\leq1.15$~\risco) regardless of the source position along the Z-track or how the 1~keV feature is modeled. Given the optically determined NS mass of $1.71\pm0.21$~\msun, this corresponds to a conservative upper limit of \rin~$\leq19.5$~km for $M=1.92$~\msun or \rin~$\leq15.3$~km for $M=1.5$~\msun. 
We compare these radius constraints to those obtained from NS gravitational wave merger events and recent \nicer pulsar light curve modeling measurements.
\end{abstract}


\section{Introduction}
Measuring neutron star masses and radii remains crucial for determining the equation of state (EoS) of ultradense, cold matter \citep{LP01}. 
Numerous observational methods have been developed for obtaining NS mass and/or radius \citep{ozel16} in order to narrow down the allowed region on the mass-radius ($M$--$R$) plane and rule out theoretical models. Notably, there have been enticing breakthroughs made via measuring the tidal deformability of NSs from the gravitational wave signature during NS-NS binaries merger events (e.g., GW170817: \citealt{abbott19}), as well as determining the compactness of millisecond pulsars through light curve modeling of modulations from hot spots on the NS surface as they rotate into and out of our line of sight (e.g., PSR J0030: \citealt{riley19, miller19}).

\begin{table*}[t!]
\caption{\source Observation Information}
\label{tab:obs}

\begin{center}
\begin{tabular}{llcccc}
\hline

Obs \# & Mission & Sequence ID & Obs.\ Start Date & Exp.\ (ks) \\
\hline
1 & NuSTAR &30001141002 & 2015-01-07 03:16:07 & $\sim23.7$ \\
2 & NuSTAR & 80511301002 & 2019-09-10 13:06:09 & $\sim11.3$\\
& NICER & 2631010101 & 2019-09-10 12:58:20 & $\sim12.7$ \\
3 & NuSTAR & 80511301004 & 2019-09-12 02:06:09 & $\sim 12.7$\\
& NICER & 2631010201 & 2019-09-12 02:09:44 & $\sim 12.1$ \\
\hline

\end{tabular}
\end{center}
\end{table*}

An additional method of independently determining NS radii can be obtained from modeling the reprocessed emission from the innermost accretion disk that has been externally illuminated. This is commonly referred to as the `reflection' spectrum which has a series of narrow emission lines superimposed on a reprocessed continuum. Emission from the inner disk region (most prominently seen in the Fe~K line) is broadened due to Doppler, general, and special relativistic effects \citep{fabian89, fabian00} which allows for a measurement of the position of the inner edge of the disk. Since the accretion disk must truncate at or prior to the NS surface, determining the inner disk radius provides a limit on the radius of the NS \citep{cackett08, ludlam17a}.

One system that has potential for demonstrating the power of NS reflection studies is the luminous, persistently accreting low-mass X-ray binary (LMXB) Cygnus X-2 (\source), especially since it has an optically determined mass. The source was first observed in the X-rays via a sounding rocket in the 1960s \citep{byram66}. \source was tentatively classified as an NS when a weak X-ray burst was observed with the {\it Einstein Observatory} \citep{khan84} and later confirmed when {\it RXTE} observed a Type-I X-ray burst while the source was in a high-intensity state \citep{smale98}. \source is classified as a `Z' source based on the tracks traced out in hardness and color-color diagrams \citep{HK89}. However, the exact shape and location on these diagrams varies depending on the overall intensity level \citep{kuulkers96, wijnands97, fridriksson15}, which can vary by a factor of $\sim4$ \citep{wijnands01}. The neutral hydrogen column density along the line of sight is low ($N_{\rm H}\sim2\times10^{21} \ {\rm cm}^{-2} $: \citealt{HI4}) with the abundance of oxygen being slightly supersolar  ($A_{\rm O}/A_{\odot}=1.1$: \citealt{psaradaki20}). 

Due to the low column density, the source has been observed extensively in the optical as well. The stellar companion is an evolved, late-type A9III star in a $9.8444\pm0.0003$ day orbit \citep{casares98}. From modeling of ellipsoidal light curves, the mass function of the system was estimated as $f(M)=0.66\pm0.03$ \msun \citep{casares10}, which leads to an estimate of the NS mass of $1.71\pm0.21$ \msun for an inclination of $62.5^{\circ} \pm4^{\circ}$ \citep{orosz99}. 
\source is estimated to be located at a distance of $8-11$~kpc \citep{cowley79, smale98}, though optical observations tend toward the lower end of this ($7.2\pm1.1$~kpc: \citealt{orosz99}).
More recently, \citet{ding21} estimated a distance of $11.3_{-0.8}^{+0.9}$~kpc using a Bayesian inference approach that utilized information from Gaia Early Data Release 3 and photospheric radius expansion bursts. 

%\begin{wrapfigure}[23]{r}[0pt]{0.48\textwidth}
\begin{figure*}[t!]
%\vspace{-7pt}
\begin{centering}
\includegraphics[width=7.9cm, trim=0 0 0 5, clip]{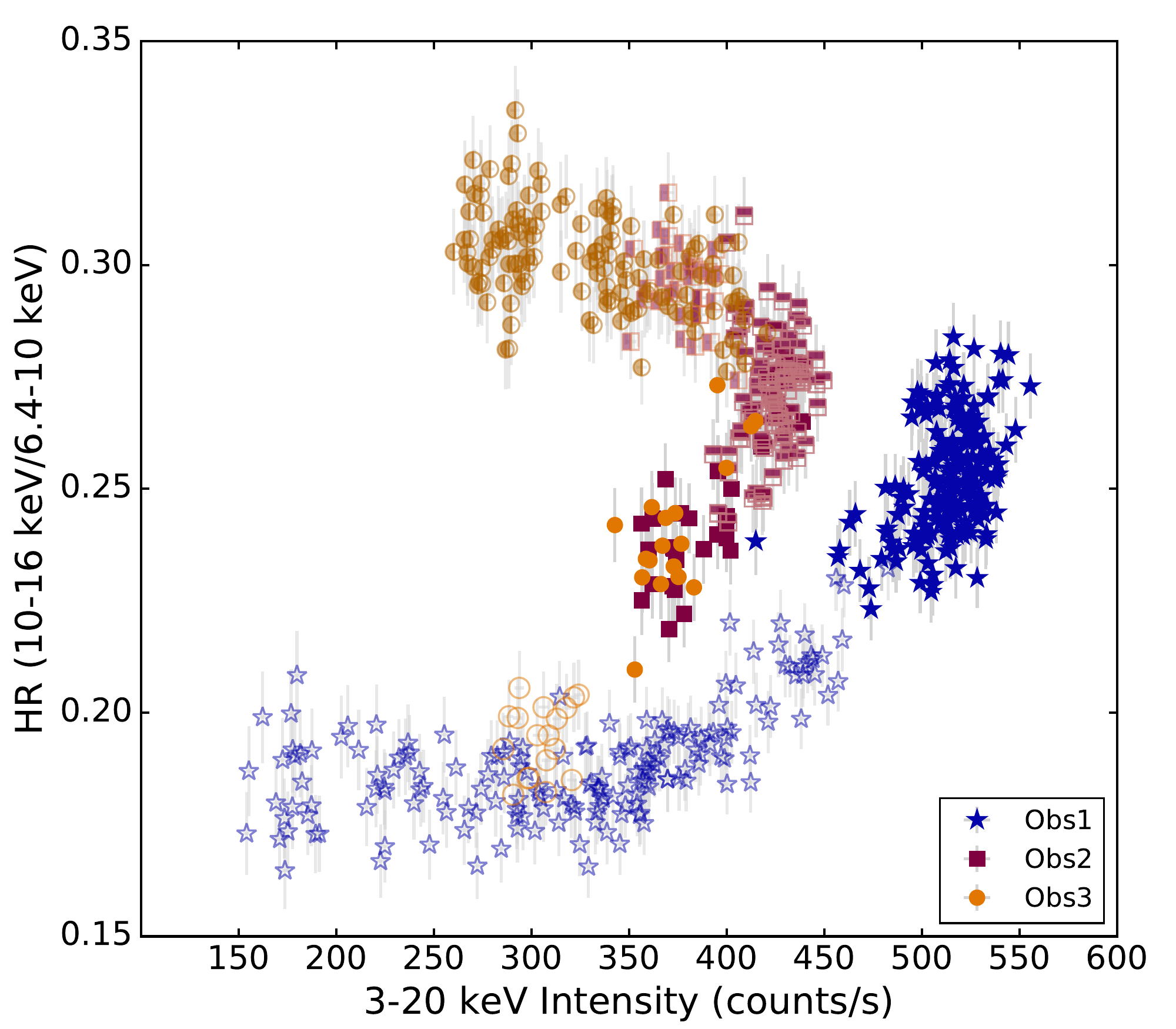}
\includegraphics[width=9.6cm, trim=0 0 20 10, clip]{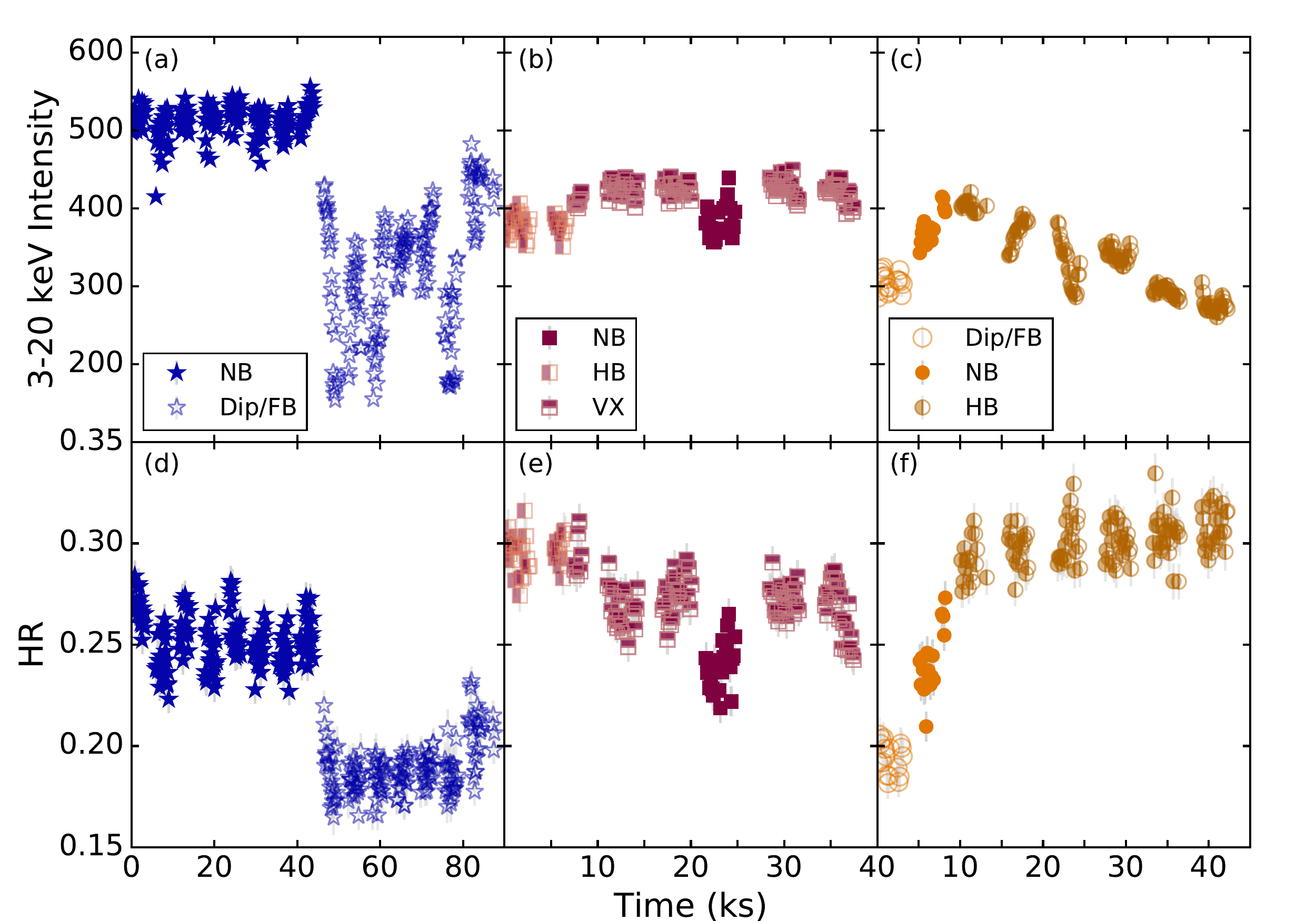}
\caption{ {\it Left:} The NuSTAR hardness ratio (HR) versus the $3-20$ keV intensity. The shape of the symbols indicate the observation. The variation in coloring and shading between symbols indicate the spectral state where solid is the normal branch (NB), top half filled is the vertex (VX), left half filled is the horizontal branch (HB), and open indicates when the source dipping or in the flaring branch (Dip/FB). {\it Right:} The top row shows the \nustar the $3-20$ keV light-curve of \source for (a) Obs1, (b) Obs2, and (c) Obs3. The lower panels show the hardness ratio (HR) during the observations. The color/shading of the symbols are coded based on the HID. Data were binned to 128~s. }
\label{fig:hid}
\end{centering}
\end{figure*} 

\begin{figure}[b!]
%\vspace{-7pt}
\begin{centering}
\includegraphics[width=8.6cm, trim=0 0 0 0, clip]{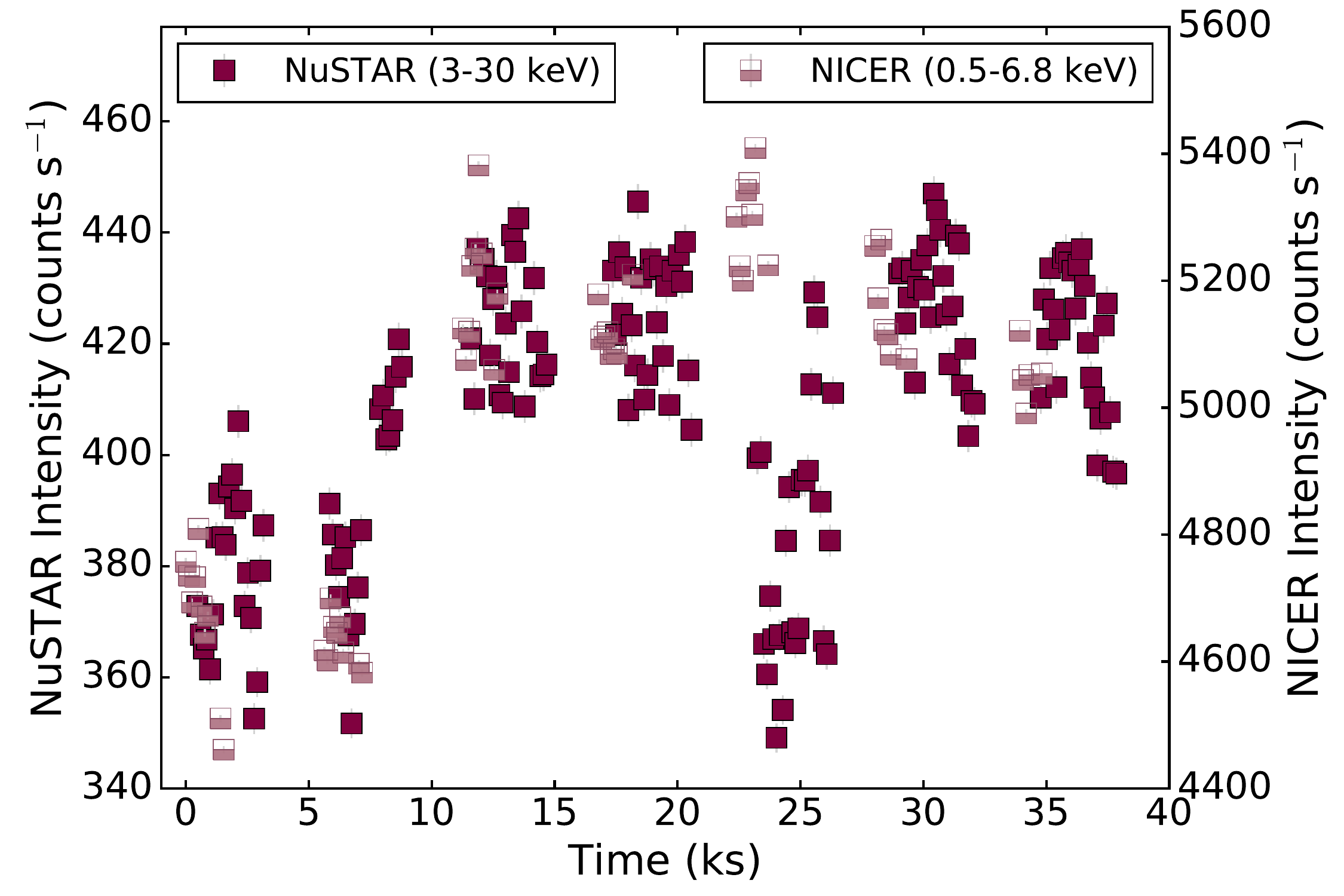}
\includegraphics[width=8.6cm, trim=0 0 0 0, clip]{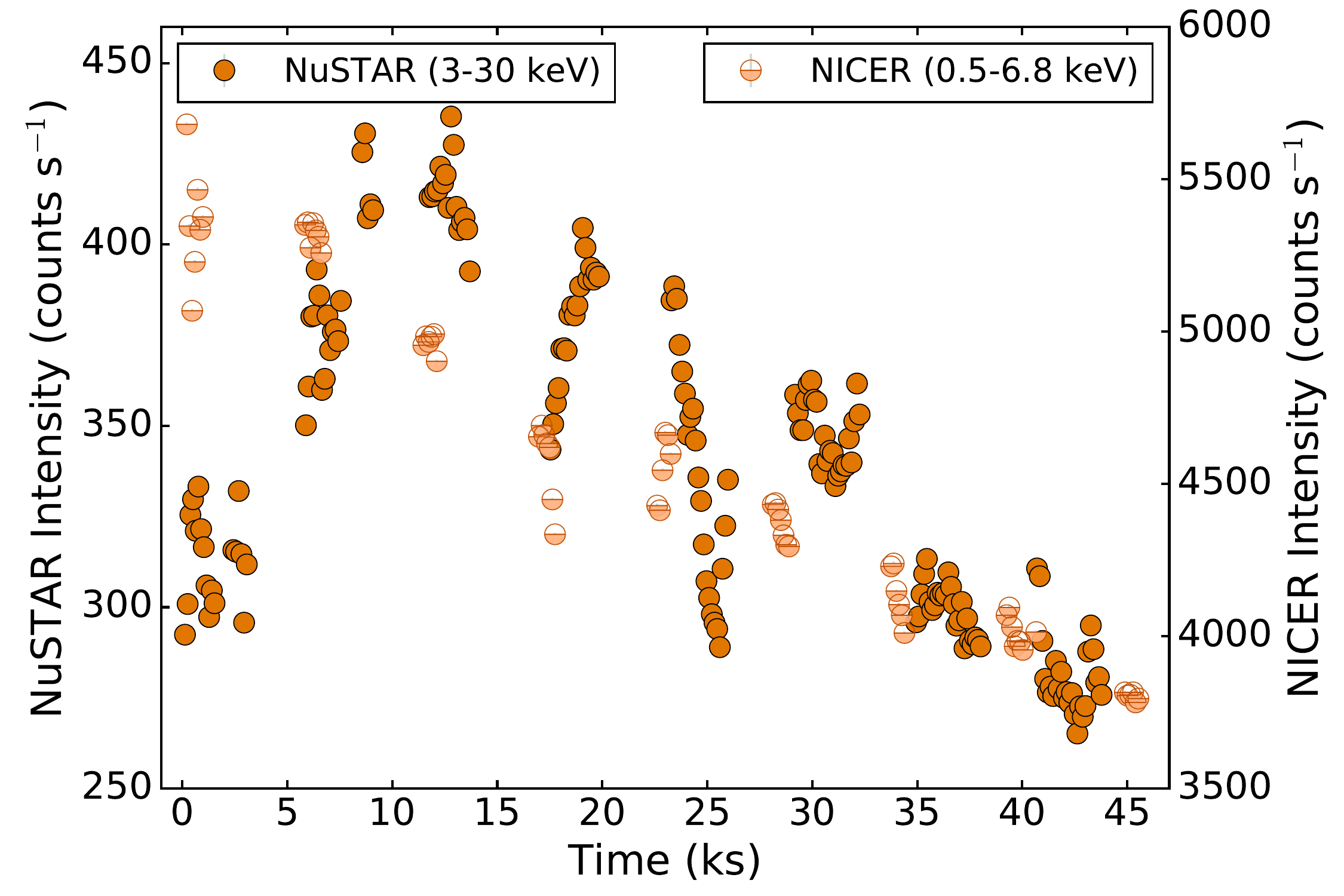}
\caption{The light curves of the simultaneous NuSTAR and NICER observations of \source for Obs2 (upper panel) and Obs3 (lower panel).  Data were binned to 128~s. }
\label{fig:ninulc}
\end{centering}
\end{figure} 

Due to the luminous and persistent nature of the source, the spectral properties have been studied considerably. The source spectrum is known to show a broad Fe line feature near 6.7~keV \citep{smale93, disalvo02, shaposhnikov09, cackett10, mondal18} due to reflection from the accretion disk, as well an emission line near 1 keV \citep{vrtilek86, chiappetti90, smale93, kuulkers97, disalvo02, farinelli09, cackett10} that likely originates from collisionally excited or photoionized material further out in the disk \citep{vrtilek86}. The source traces out the horizontal, normal, and flaring branches of the `Z' with periods of irregular dipping activity while flaring; suggesting the presence of an extended accretion disk corona (ADC) during high intensity states \citep{vrtilek88, schulz09, bc10, bc11}. A detailed broadband spectral analysis of \source with {\it BeppoSAX} while the source was in the horizontal and normal branches was reported in \citet{disalvo02}. The continuum modeling suggested that the inner accretion disk moved closer to the NS as the inferred mass accretion rate increased as the source transitioned from the horizontal to the normal branch, but a full treatment of the reflection spectrum was not conducted.

Reflection modeling of {\it Suzaku} observations was performed in \citet{cackett10} using a {\sc diskline} component and blackbody reflection model ({\sc bbrefl}) to obtain inner disk radius and inclination constraints. The disk was inferred to be close to the NS at $R_{\rm in}\simeq7.6-8.5 \ R_{g}$ (where $R_{g}= GM/c^{2}$) when using single {\sc diskline} component and $R_{\rm in}\simeq6-14 \ R_{g}$ when using the full reflection model. 
\citet{mondal18} recently analyzed a \nustar observation of \source in the normal branch and flaring/dipping state. The reflection component is modeled with a different blackbody reflection model known as {\sc reflionxbb} and the disk was inferred to be far from the NS at $R_{in}\simeq13.5-32.4\ R_{g}$ in the non-dipping state. The inferred inclination in both of these studies ($i\lesssim25^{\circ}$), however, are at odds with the inclination measured from optical observations. This may be due to the thickness of the disk in the outer regions being able to partially obscure the blue-winged emission of the Fe~K line \citep{taylor18}. 

Here we analyze the existing NuSTAR observations of \source while the source is in a non-dipping state in order to carefully obtain radius constraints from reflection modeling. Though one \nustar observation was reported on in \citet{mondal18}, two additional \nustar observations were performed simultaneously with \nicer. Hence we present the results of joint \nicer and \nustar spectral modeling. The organization of the paper is as follows: \S 2 presents the observations and data reduction methodology, \S 3 reports the spectral modeling and results, \S 4 discusses the results and compares them to the current best constraints on NS mass and radius, \S 5 then provides the conclusion.

\section{Observations and Data Reduction}

\nustar has observed \source on three occasions. The sequence IDs, observation dates, and exposure time are given in Table \ref{tab:obs}. All data were reduced using CALDB v.20210427 and `nupipeline' with `statusexpr``STATUS==b0000xxx00xxxx000"' due to the source brightness having an excess of 100 \cps. The background filtering report for Obs1 indicated periods of high background, hence we applied ``saacalc=3 saamode=optimized tentacle=yes". Source regions of $100''$ radii centered on the source and a background region of same size but sufficiently far from the source were used for spectra and light curve extraction. The light curves were inspected for Type-I X-ray bursts, but none were present. The hardness ratio (HR: $10-16$ keV/$6.4-10$ keV) versus $3-20$ keV intensity, known as the  hardness-intensity diagram (HID), is shown in Figure \ref{fig:hid}, as well as the light curves and hardness ratio versus time for each \nustar observation. The observations trace out the flaring to the horizontal branch in the HID and shows shifts in the overall intensity between the 2015 and 2019 observations. 

\nicer observed the source simultaneous with \nustar during Obs2 and Obs3. Information for each sequence ID is given in Table \ref{tab:obs}. Data were reduced and calibrated using the standard `nicerl2' command and CALDB version 20200722. Additionally, the data were filtered to select for $\rm{KP}<5$ and $\rm{COR\_SAX}>4$ to mitigate the particle background at low-energies. The simultaneous light curves for these observations are shown in Figure \ref{fig:ninulc}, where the zero point indicates the start of the \nicer observation for Obs2 and the beginning of the \nustar observation for Obs3. No Type-I X-ray bursts were present in the \nicer data.

\movetableright=-15mm
\begin{table*}[t!]
\caption{Continuum Spectral Modeling}
\label{tab:continuum} 

\begin{center}
\begin{tabular}{llccc|ccc}
\hline

%VX_NB_HB_crabcor_tbfeo_diskbbodypow_ninu_chi_06282021_5E4_burn1E6.out
%VX_NB_HB_crabcor_tbfeo_disknthcomp_ninu_chi_06282021_5E4_burn1E6.out

Model & Parameter & \multicolumn{3}{c}{C1} & \multicolumn{3}{c}{C2}\\
& & NB & VX & HB & NB & VX & HB\\
\hline

{\sc crabcor} %%%%
& $C_{\rm FPMB}$ 
&$ 1.022\pm0.001$
&$ 1.014 _{- 0.002 }^{+ 0.001 }$
& $ 1.013 _{- 0.001 }^{+ 0.002 }$

&$ 1.013 _{- 0.001 }^{+ 0.003 }$
&$ 1.014 _{- 0.001 }^{+ 0.002 }$
&$ 1.022 _{- 0.001 }^{+ 0.002 }$
\\
& $C_{\rm NICER}$
& ...
& $ 0.99\pm0.01$
&$ 1.02 \pm0.01$
& ...
& $ 1.01\pm0.01$
& $ 0.98\pm0.01$
\\
& $\Delta \Gamma_{\rm NICER}\ ^{b}$ ($10^{-2}$)
& ...
& \multicolumn{2}{c|}{$ -4.4 _{- 0.4 }^{+ 0.6 }$}
& ...
& \multicolumn{2}{c}{ $ -5.1 _{- 0.4 }^{+ 0.7 }$}
\\

{\sc tbfeo} %%%%
&$\mathrm{N}_{\mathrm{H}}\ ^{a}$ ($10^{21}$ cm$^{-2}$) 
& ---
&$ 4.19 _{- 0.06 }^{+ 0.13 }$
&---
&---
& $ 2.2 _{- 0.2 }^{+ 0.1 }$
&---
\\
&$A_{\rm O}\ ^{a}$
&---
&$ 1.09 _{- 0.03 }^{+ 0.04 }$
&---
&---
&$ 1.32 _{- 0.02 }^{+ 0.07 }$
&---
\\

{\sc diskbb} %%%%
& $kT_{\rm in}$ (keV) 
&$ 1.76\pm0.01$
&$ 1.80 \pm0.01$
&$ 1.78\pm0.01$
&$ 1.72 \pm0.01$
&$ 1.80 \pm0.01$
&$ 1.83 _{- 0.01 }^{+ 0.02 }$
\\

& norm$_{\rm disk}$ 
&$ 119 _{- 2 }^{+4 }$
&$ 83 _{- 1}^{+ 2 }$
&$ 60 \pm1$
&$ 110 _{- 6}^{+ 2}$
&$ 69 \pm2$
&$ 41 _{- 2}^{+ 1}$
\\

{\sc bbody} %%%%
& $kT_{\rm bb}$ (keV) 
&$ 2.68 _{- 0.03 }^{+ 0.01 }$
&$ 2.72 _{- 0.02 }^{+ 0.03 }$
&$ 2.67 \pm0.02$
& ...
& ...
& ...
\\
& norm$_{\rm bb}$ ($10^{-2}$)
& $ 3.37 _{- 0.07 }^{+ 0.14 }$
& $ 3.6 \pm0.1$
& $ 3.71 _{- 0.09 }^{+ 0.07 }$
& ...
& ...
& ...
\\

{\sc powerlaw} %%%%
& $\Gamma$ 
&  $ 3.96 _{- 0.09 }^{+ 0.04 }$
& $ 3.19 _{- 0.01 }^{+ 0.11 }$
& $ 2.99 _{- 0.04 }^{+ 0.06 }$
& ...
& ...
& ...
\\
& norm$_{\rm pl}$ 
&$ 7\pm1$
&$ 1.5 \pm0.1$
&$ 1.6 \pm0.1$
& ...
& ...
& ...
\\

{\sc nthcomp} %%%%
& $\Gamma$
& ...
& ...
& ...
&$ 1.77 _{- 0.02 }^{+ 0.06 }$
&$ 1.69 _{- 0.02 }^{+ 0.03 }$
&$ 1.78 _{- 0.01}^{+ 0.02 }$
\\

& $kT_{e}$ (keV)
& ...
& ...
& ...
& $ 2.89 _{- 0.02 }^{+ 0.05 }$
&$ 3.03 _{- 0.04 }^{+ 0.03 }$
& $ 3.11 _{- 0.01 }^{+ 0.05 }$
\\

& $kT_{bb}$ ($10^{-1}$~keV) 
& ...
& ...
& ...
&$ 1.27 _{- 1.24 }^{+ 0.07}$
&$ 1.6 _{- 0.4 }^{+ 0.1 }$
& $ 1.8 _{- 0.25 }^{+ 0.08 }$
\\

& norm$_{\rm nth}$ 
& ...
& ...
& ...
&$ 1.35 _{- 0.07 }^{+ 0.27 }$
&$ 0.88 _{- 0.05 }^{+ 0.10 }$
&$ 1.19 _{- 0.04 }^{+ 0.07 }$
\\

& $F_{\rm unabs,\ 0.5-50\ keV}$ 
&$5.0\pm0.3$
&$2.6\pm0.2$
&$2.1\pm0.3$
&$3.0\pm0.2$
&$2.3_{-0.1}^{+0.3}$
&$1.9_{-0.1}^{+0.4}$
\\

\hline
& $\chi^{2}$ (dof) & \multicolumn{3}{c}{ 3380.0 (1952)} & \multicolumn{3}{c}{3386.8 (1952)}\\ 
\hline

\multicolumn{6}{l}{$^{a}=$ tied between all branches, $^{b}=$ tied between NICER spectra} 

\end{tabular}
\end{center}

\medskip
Note.---  Errors are reported at the 90\% confidence level. \nicer\ is fit in the $0.5-10$ keV energy band while \nustar is fit in the $3-30$ keV band.  A multiplicative constant is used on the \nicer and FPMB data, while FPMA is fixed to unity. The input seed photon type in {\sc nthcomp} is set to a single temperature blackbody (inp\_type=0). 
The {\sc bbody} normalization is defined as $(L/10^{39}\ \mathrm{erg\ s^{-1}})/(D/10\ \mathrm{kpc})^{2}$. 
The {\sc diskbb} normalization is defined as $(R_{in}/\mathrm{km})^{2}/(D/10\ \mathrm{kpc})^{2}\times\cos{\theta}$. 
The power-law normalization is defined as photons~keV$^{-1}$~cm$^{-2}$~s$^{-1}$ at 1~keV.
The unabsorbed $0.5-50$~keV flux, $F_{\rm unabs,\ 0.5-50\ keV}$, is given in units of $10^{-8}$ \fluxcgs. 

\end{table*}

The \nustar and \nicer data were divided into the different branches within each observation using good time intervals (GTIs) based on the HID and light curves shown in Figures \ref{fig:hid} and \ref{fig:ninulc}. 
The source went through the normal to extended flaring branch in Obs1 while in the high intensity state as reported in \citet{mondal18}. We divide the data in a similar manner by separating the non-dip emission from the dipping by creating GTIs that divide the observation at t~$=45$~ks. Obs2 and Obs3 occurred while the source was in a lower intensity state. Obs2 traced out the upper normal branch (NB), vertex (VX), and horizontal branch (HB). The HB occurred from t~$<7.5$~ks, NB from t~$=21.5-27.5$~ks, and the VX from t~$=7.5-21.5$~ks and t~$>27.5$~ks. Obs3 occurred as the source was exiting a dipping period. Therefore, the first 5 ks were removed. The remaining data were divided between the NB (t~$=5-10$~ks) and HB (t~$>10$~ks). These GTIs were applied to the \nustar data through `nuproducts' to extract spectra while the data were in the NB, HB, and VX. The \nicer data were divided into these different branches by converting the \nustar GTIs into \nicer mission elapsed time (MET) and then extracting the corresponding events through `niextract-events'. Source and background spectra for \nicer data were created using the `3C50' tool\footnote{https://heasarc.gsfc.nasa.gov/docs/nicer/tools/nicer\_bkg\_est\_tools.html} \citep{remillard21}.

\begin{figure*}[!]
%\vspace{-7pt}
\begin{centering}
\includegraphics[width=17.5cm, trim=10 0 0 0, clip]{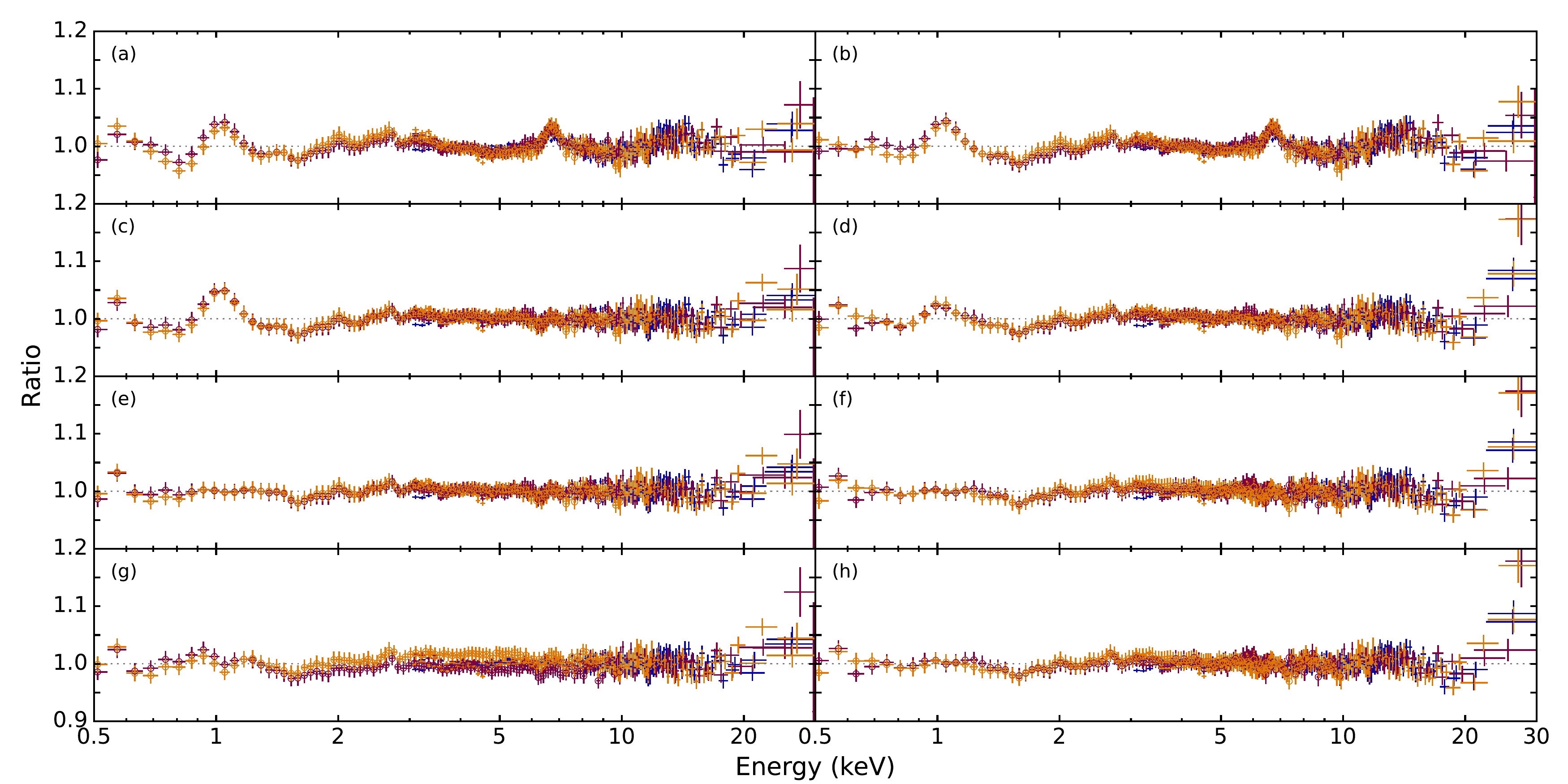}
\includegraphics[width=17.6cm, trim=2 0 0 0, clip]{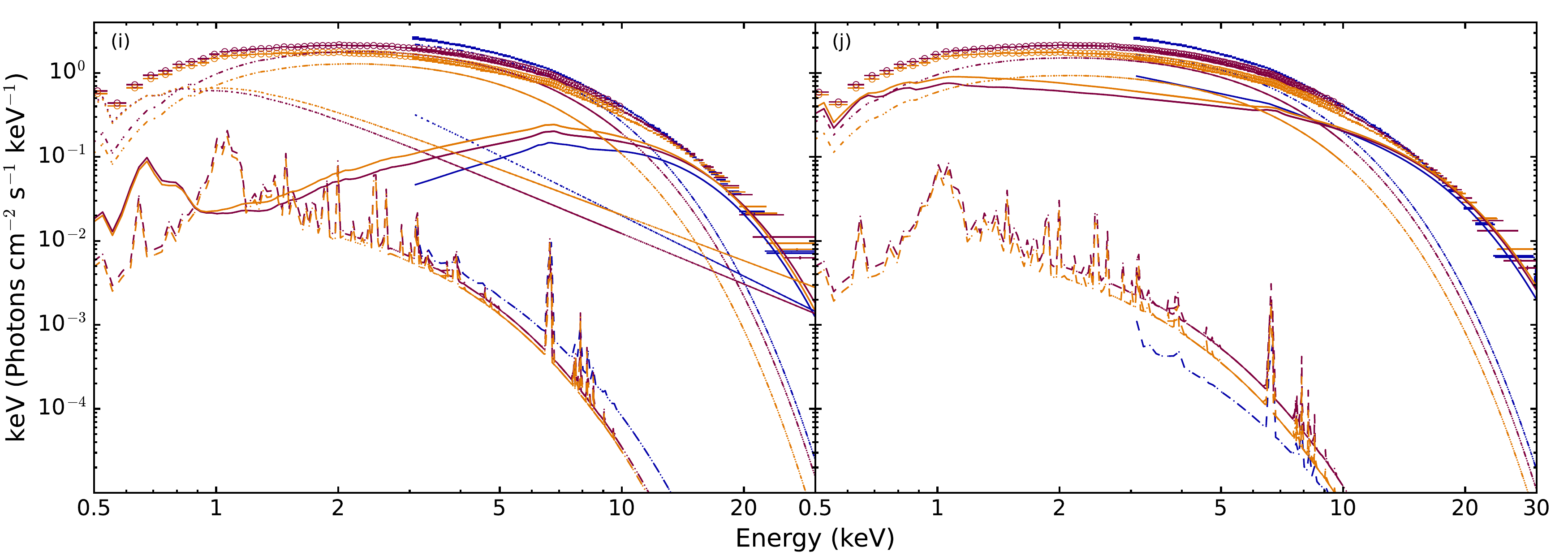}
\caption{Ratio of the data to the models reported in Tables \ref{tab:continuum}-\ref{tab:nthrefl}, where panels correspond to (a) C1, (b) C2, (c) RNS1, (d) RFX1, (e) RNS2, (f) RFX2, (g) RNS3, and (h) RFX3. Blue indicates the NB from Obs1, maroon is the VX from Obs2, and orange is the HB from Obs3. Data were rebinned for plotting purposes. Panels (i) and (j) show the unfolded spectra and model components for RNS3 and RFX3, respectively. The dot-dashed line is the disk component, dotted line is the power law component, dashed line is the collisional plasma, and the solid line indicates the reflection model with the corresponding illuminating component included. }
\label{fig:ratios}
\end{centering}
\end{figure*}

\section{Spectral Modeling and Results}
We utilize {\sc xspec} v12.11.1 to model all spectra simultaneously. To account for the cross calibration difference between NICER and NuSTAR, we use the {\sc crabcorr}  multiplicative model (a.k.a. {\sc jscrab}; \citealt{steiner10}). This model has two parameters: 1.) $\Delta \Gamma$ that multiplies the spectrum by a power law difference (E$^{-\Delta \Gamma}$) and 2.) a normalization, $C$, that serves in the same capacity as a multiplicative constant. $\Delta \Gamma$ is set to 0 for the \nustar spectra and allowed to float for the \nicer. Although the value of $\Delta \Gamma$ is small, it is important to account for unavoidable mission-specific calibration differences that emerge when the Crab is observed.
The multiplicative constant was allowed to vary for the FPMB and NICER while the FPMA was fixed at unity. 
The absorption column along the line of sight was modeled with {\sc tbfeo} with abundances set to {\sc wilm} \citep{wilms00} and {\sc vern} \citep{verner96} cross-sections. The column density, $N_{\rm H}$, and abundance of oxygen, $A_{\rm O}$, were allowed to vary but tied between all spectra regardless of spectral state. Errors are reported at the 90\% confidence level from Markov Chain Monte Carlo (MCMC) with 50 walkers, a burn-in of $1\times10^6$, and chain length of $5\times10^4$. 

In the interest of obtaining robust constraints on the inner disk radius, we focus on modeling the non-dipping spectra that have $\geq10^{6}$ cumulative counts per spectrum. This corresponds to the \nustar spectra of Obs1 in the NB, \nicer and \nustar spectra of Obs2 in the VX, and \nicer and \nustar spectra of Obs3 in the HB. These will be referred to by their branch nomenclature in Tables \ref{tab:continuum} -- \ref{tab:nthrefl}. We model the continuum with the phenomenological three component model of \citet{lin07}. This model is comprised of a multi-temperature blackbody to account for the thermal disk emission ({\sc diskbb}: \citealt{mitsuda94}), a single temperature blackbody ({\sc bbody}) for emission from the NS surface or boundary layer region, and a power law ({\sc pow}) component to model weak Comptonization. This continuum model is referred to as ``C1" in Table \ref{tab:continuum}. The ratio of the model to the data are shown in Figure \ref{fig:ratios}(a). Additionally, we swap out the power law and single-temperature blackbody component for a more physical Comptonization model, {\sc nthcomp} \citep{zdziarski96,zycki99}. Parameter values can be found in Table \ref{tab:continuum} under ``C2" and the ratio of the model to the data are shown in Figure \ref{fig:ratios}(b).  The fits are of comparable statistical quality, though C1 provides a slightly better fit. The spectral parameters for the NB are relatively consistent with those reported in \citet{mondal18}, but it is important to note that differences likely arise from the bright source flag being utilized when we reduced the \nustar data. 

\begin{figure}[!]
%\vspace{-7pt}
\begin{centering}
\includegraphics[width=8.2cm, trim=0 0 0 10, clip]{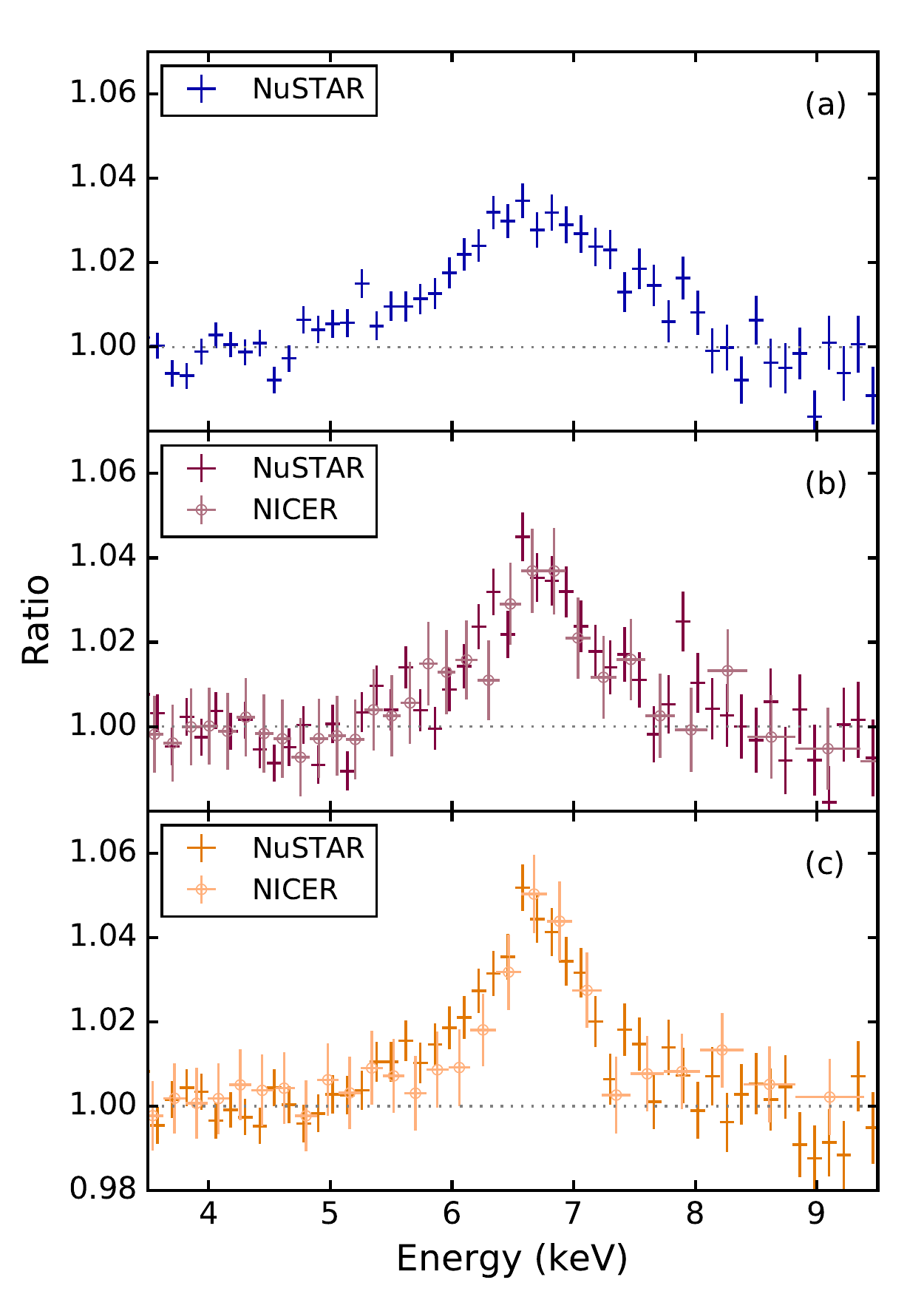}
\caption{Ratio of the data in the Fe K band to the continuum model C1 for the (a) NB, (b) VX, and (c) HB. Only one \nustar FPM is shown for clarity. Data were rebinned for plotting purposes. }
\label{fig:Fe}
\end{centering}
\end{figure} 

There is clear evidence of a broadened Fe~K line component in all spectra regardless of continuum model. Figure~\ref{fig:Fe} shows the Fe line profile in each respective branch. 
Additionally, there is an emission line present in the \nicer spectra at lower-energy ($\sim1$~keV). As mentioned in the introduction, this feature has been reported previously with other X-ray missions. 

We model the reflected emission in two separate ways corresponding to the input continuum. Starting from C1, we utilize the variation of \relxill \citep{garcia14} known as \relxillns to account for reprocessed emission due to an illuminating blackbody component, $kT_{bb}$. The model parameters are as follows: an inner emissivity index ($q_{in}$), outer emissivity index ($q_{out}$), the break radius ($R_{break}$) between the two emissivity indices, dimensionless spin parameter ($a$), redshift ($z$), inclination of the system ($i$), inner disk radius (\rin) in units of the innermost stable circular orbit (ISCO), outer disk radius ($R_{out}$) in units of gravitational radii ($R_{g}$), ionization parameter ($\log(\xi)$), disk density ($\log(N \mathrm{[cm^{-3}]})$), iron abundance ($A_{Fe}$), and reflection fraction ($f_{refl}$). We tie the inner and outer emissivity indices to create a single emissivity profile, $q$, and therefore $R_{break}$ becomes irrelevant. \source\ is a Galactic source so we fix the redshift parameter to $z=0$. The outer disk radius is set at 1000 \rg\ and the spin parameter is fixed at $a=0$.
The choice of spin parameter is motivated by most NSs in LMXBs having $a\lesssim0.3$ \citep{galloway08, miller11} and the difference in the position of \risco between these two values being less than 1~\rg \citep{ludlam18}. 
The reflection fraction, $f_{refl}$, is bound to positive values so that the \relxillns model encompasses both the illuminating blackbody from the continuum and reprocessed reflection emission component. These model fits are reported in Table~\ref{tab:bbrefl} under ``RNS1".

\movetabledown=55mm
%\hspace{-50mm}

\begin{table*}
\begin{rotatetable*}
\caption{Reflection Modeling from a Single-temperature Blackbody}
\label{tab:bbrefl}

\begin{center}

\begin{tabular}{llccc|ccc|ccc}
\hline

%VX_NB_HB_crabcor_tbfeo_diskpowrelxillns_ninu_chi_06292021_5E4_burn1E6.out
%VX_NB_HB_crabcor_tbfeo_diskpowrelxillns_gauss_ninu_chi_07022021_5E4_burn1E6.out
%VX_NB_HB_crabcor_tbfeo_diskpowrelxillns_mekal_ninu_chi_06292021_5E4_burn1E6.out

Model & Parameter & \multicolumn{3}{c}{RNS1} & \multicolumn{3}{c}{RNS2} & \multicolumn{3}{c}{RNS3}\\
&& NB & VX & HB & NB & VX & HB & NB & VX & HB\\
\hline

{\sc crabcor} %%%%
& $C_{\rm FPMB}$
&$ 1.022 \pm0.001$
&$ 1.014 \pm0.002$
&$ 1.014 _{- 0.001 }^{+ 0.003 }$
&$ 1.022 _{- 0.001 }^{+ 0.002 }$
&$ 1.014\pm 0.002$
&$ 1.015 _{- 0.003 }^{+ 0.001 }$
&$ 1.023 _{- 0.002 }^{+ 0.001}$
&$ 1.014\pm 0.002$
&$ 1.015 _{- 0.002 }^{+ 0.001 }$
\\

& $C_{\rm NICER}$ 
&...
&$ 0.969 _{- 0.008 }^{+ 0.006 }$
&$ 0.996 _{- 0.007 }^{+ 0.005 }$
&...
&$ 0.974 _{- 0.002 }^{+ 0.007 }$
&$ 1.000 _{- 0.002 }^{+ 0.007 }$
&...
& $ 0.966 _{- 0.007 }^{+ 0.004 }$
&$ 0.993 _{- 0.008 }^{+ 0.005 }$
\\

& $\Delta \Gamma_{\rm NICER} \ ^{b}$ ($10^{-2}$)
&...
& \multicolumn{2}{c|}{ $ -5.7\pm0.4$}
&...
& \multicolumn{2}{c|}{$ -5.7 \pm0.2$}
&...
& \multicolumn{2}{c}{$ -5.8 _{- 0.6 }^{+ 0.2 }$}
\\

{\sc tbfeo} %%%%
& $\mathrm{N}_{\mathrm{H}}\ ^{a}$ ($10^{21}$ cm$^{-2}$) 
& ---
& $ 3.92 _{- 0.07 }^{+ 0.08 }$
& ---
& ---
&$ 3.71 _{- 0.05}^{+ 0.09 }$
& ---
& ---
& $ 3.65 _{- 0.07 }^{+ 0.14 }$
& ---
\\
& $A_{\rm O}\ ^{a}$
& ---
& $ 1.16 _{- 0.01 }^{+ 0.05 }$
& ---
& ---
&$ 1.17 _{- 0.03 }^{+ 0.04 }$
& ---
& ---
& $ 1.19 _{- 0.06 }^{+ 0.03 }$
& ---
\\

{\sc diskbb} %%%%
& $kT_{\rm in}$ (keV) 
&$ 1.77 _{- 0.01 }^{+ 0.02 }$
&$ 1.73 _{- 0.01 }^{+ 0.03 }$
& $ 1.65 _{- 0.03 }^{+ 0.01 }$
&$ 1.77 \pm0.01$
&$ 1.73 \pm0.02$
&$ 1.61 _{- 0.01 }^{+ 0.04 }$
&$ 1.78 _{- 0.02 }^{+ 0.01 }$
&$ 1.74 \pm0.02$
&$ 1.63 _{- 0.01 }^{+ 0.02 }$
\\

& norm$_{\rm disk}$ 
&$ 114 _{- 4 }^{+ 2 }$
&$ 94 _{- 5 }^{+ 1 }$
&$ 78 _{- 2 }^{+ 4 }$
&$ 116 _{- 3 }^{+ 2 }$
&$ 94 _{- 4 }^{+ 3 }$
&$ 86 _{- 7 }^{+ 1}$
&$ 114 _{- 3 }^{+ 4 }$
&$ 91 _{- 2 }^{+ 4 }$
&$ 82 _{- 4 }^{+ 2}$
\\

{\sc powerlaw} %%%%
& $\Gamma$ 
&$ 3.43 _{- 0.10 }^{+ 0.08 }$
& $ 3.02\pm0.05$
& $2.84 _{- 0.04 }^{+ 0.03}$
& $ 3.41 _{- 0.10 }^{+ 0.01 }$
&$ 2.95 \pm0.06$
&$ 2.79 \pm0.03$
&$ 3.44 _{- 0.10 }^{+ 0.01 }$
&$ 2.94 _{- 0.04 }^{+ 0.09 }$
& $ 2.79 _{- 0.03 }^{+ 0.05 }$
\\
& norm$_{\rm pl}$ 
&$ 5.2 _{- 0.4 }^{+ 0.7 }$
&$ 1.41 _{- 0.05 }^{+ 0.09 }$
& $ 1.49 _{- 0.03 }^{+ 0.08 }$
&$ 4.9 _{- 0.3}^{+ 0.2}$
&$ 1.19 \pm0.07$
&$ 1.26 _{- 0.04 }^{+ 0.10 }$
&$ 4.7 _{- 0.2 }^{+ 0.3 }$
&$ 1.18 _{- 0.11 }^{+ 0.08 }$
&$ 1.23 _{- 0.05 }^{+ 0.12}$
\\

{\sc relxillNS} %%%%
& $q$
&$ 2.19 _{- 0.09 }^{+ 0.10 }$
&$ 1.8\pm0.1$
&$ 1.8\pm0.2$
&$ 2.2 _{- 0.2 }^{+ 0.1 }$
&$ 1.9 _{- 0.2 }^{+ 0.1 }$
&$ 1.8 _{- 0.2 }^{+ 0.1 }$
&$ 2.2 \pm0.1$
&$ 1.8 _{- 0.1 }^{+ 0.2 }$
&$ 1.7 _{- 0.1 }^{+ 0.2 }$
\\
& $i \ ^{a}$ ($^{\circ}$)
& ---
&$ 68 _{- 5 }^{+ 1 }$
& ---
&---
&$ 67 _{- 3 }^{+ 4 }$
&---
&---
&$ 67 _{- 4 }^{+ 2 }$
&---
\\
& \rin\ (ISCO)
&$ 1.01 _{- 0.01 }^{+ 0.06 }$
&$ 1.06 _{- 0.06 }^{+ 0.09 }$
&$ 1.07 _{- 0.07 }^{+ 0.03 }$
&$ 1.01 _{- 0.01 }^{+ 0.14 }$
&$ 1.05 _{- 0.05 }^{+ 0.11 }$
&$ 1.11 _{- 0.11 }^{+ 0.04 }$
&$ 1.02 _{- 0.02 }^{+ 0.06 }$
&$ 1.09 _{- 0.09 }^{+ 0.06 }$
&$ 1.03 _{- 0.03 }^{+ 0.08 }$
\\
& \rin\ (\rg)
&$ 6.06 _{- 0.06}^{+ 0.36 }$
&$ 6.36 _{- 0.36 }^{+ 0.54 }$
&$ 6.42 _{- 0.42 }^{+ 0.18 }$
&$ 6.06 _{- 0.06 }^{+ 0.84 }$
&$ 6.30 _{- 0.30 }^{+ 0.66 }$
&$ 6.66 _{- 0.66 }^{+ 0.24 }$
&$ 6.12 _{- 0.12 }^{+ 0.36 }$
&$ 6.54 _{- 0.54 }^{+ 0.36 }$
&$ 6.18 _{- 0.18 }^{+ 0.48 }$
\\
& $kT_{bb}$ (keV)
&$ 2.47 _{- 0.04 }^{+ 0.07 }$
&$ 2.56 _{- 0.03 }^{+ 0.02 }$
& $ 2.50 _{- 0.05 }^{+ 0.01 }$
&$ 2.46 _{- 0.08 }^{+ 0.04 }$
&$ 2.57 _{- 0.03 }^{+ 0.02 }$
&$ 2.46 _{- 0.01 }^{+ 0.04 }$
&$ 2.50 _{- 0.11 }^{+ 0.01}$
&$ 2.58 _{- 0.04 }^{+ 0.03 }$
& $ 2.48 _{- 0.02}^{+ 0.04 }$
\\
& $\log(\xi)$
&$ 1.51 _{- 0.05 }^{+ 0.07 }$
&$ 2.50 \pm0.03$
&$ 2.58 _{- 0.05 }^{+ 0.02}$
&$ 1.50 _{- 0.05 }^{+ 0.07 }$
&$2.50 _{- 0.05 }^{+ 0.06 }$
&$ 2.57 _{- 0.03 }^{+ 0.07 }$
&$ 1.53 _{- 0.07 }^{+ 0.09 }$
&$ 2.50 _{- 0.06 }^{+ 0.01 }$
&$ 2.57 \pm 0.04$
\\
&$A_{\rm Fe} \ ^{a}$
&---
&$ 1.34 _{- 0.01 }^{+ 0.31 }$
&---
&---
&$ 1.4 \pm0.1$
&---
&---
&$ 1.4 \pm0.1$
&---
\\
& $\log(N) \ ^{b}$ (cm$^{-3}$)
&$ 17.99 _{- 0.18 }^{+ 0.02 }$
&\multicolumn{2}{c|}{$ 17.9 _{- 0.2 }^{+ 0.1 }$}
&$ 18.0 _{- 0.2 }^{+ 0.1 }$
&\multicolumn{2}{c|}{$ 17.82 _{- 0.08 }^{+ 0.10 }$}
&$ 17.99 _{- 0.13 }^{+ 0.02 }$
&\multicolumn{2}{c}{$ 17.88 _{- 0.24 }^{+ 0.06 }$}

\\
& $f_{refl}$
&$ 0.9 _{- 0.2 }^{+ 0.1 }$
&$ 0.18 _{- 0.01 }^{+ 0.03 }$
&$ 0.13 \pm0.02$
&$ 0.8 \pm0.1$
&$ 0.19 _{- 0.02 }^{+ 0.01 }$
&$ 0.13 \pm0.01$
&$ 0.8 \pm0.1$
&$ 0.18 _{- 0.03 }^{+ 0.02 }$
&$ 0.12 _{- 0.02 }^{+ 0.01}$
\\
& norm$_{rel}$ ($10^{-3}$)
&$ 1.6 _{- 0.2 }^{+ 0.1 }$
&$ 3.2 \pm0.2$
&$ 3.7 _{- 0.1 }^{+ 0.2 }$
&$ 1.72 _{- 0.23 }^{+ 0.03 }$
&$ 3.1_{- 0.1 }^{+ 0.2 }$
&$ 3.87 _{- 0.24 }^{+ 0.04 }$
&$ 1.6 _{- 0.2 }^{+ 0.1 }$
&$ 3.2 _{- 0.1 }^{+ 0.2 }$
&$ 3.8_{- 0.2 }^{+ 0.1 }$
\\

{\sc Gaussian} %%%%
& E (keV)
&...
&...
&...
&...
&$ 1.01 _{- 0.04 }^{+ 0.02 }$
&$ 1.03 \pm0.03$
&...
&...
&...
\\
& $\sigma$ ($10^{-2}$ keV)
&...
&...
&...
&...
&$ 8.5 _{- 0.3 }^{+ 0.8 }$
&$ 8.1 _{- 0.7 }^{+ 1.3 }$
&...
&...
&...
\\
& norm$_{\rm gauss}$ ($10^{-2}$)
&....
&...
&...
&...
&$ 6.2 _{- 0.8 }^{+ 0.4 }$
&$ 4.0 _{- 0.5 }^{+ 0.3 }$
&...
&...
&...
\\
& EW (eV)
&...
&...
&...
&...
&$17_{-2}^{+1}$
&$15_{-2}^{+1}$
&...
&...
&...
\\
{\sc mekal} %%%%
& $kT$ (keV)
&...
&...
&...
&...
&...
&...
&$ 1.54 _{- 0.06 }^{+ 0.19 }$
&$ 1.37 _{- 0.09 }^{+ 0.11 }$
& $ 1.36 _{- 0.09 }^{+ 0.13 }$
\\
& A$_{\rm Fe}\ ^{a}$
&...
&...
&...
&...
&...
&...
&---
&$ 2.9_{- 0.3 }^{+ 0.4 }$
&---
\\
& norm$_{\rm mekal}$ ($10^{-2}$)
&...
&...
&...
&...
&...
&...
&$ 10_{- 2 }^{+ 1 }$
&$ 11.0 _{- 1.3 }^{+ 0.8 }$
&$ 9.1 _{- 0.9 }^{+ 1.3 }$
\\

&$F_{\rm unabs,\ 0.5-50\ keV}$
&$4.1\pm0.6$
&$2.5\pm0.2$
&$2.1_{-0.1}^{+0.2}$
&$4.0_{-0.6}^{+0.2}$
&$2.5_{-0.4}^{+0.3}$
&$2.1_{-0.3}^{+0.2}$
&$4.1_{-1.0}^{+0.6}$
&$2.5_{-0.4}^{+0.3}$
&$2.0_{-0.3}^{+0.4}$
\\

\hline
&$\chi^{2}$ (dof) & \multicolumn{3}{c}{1993.7 (1936)} & \multicolumn{3}{c}{ 1906.8 (1930)} & \multicolumn{3}{c}{ 1914.2 (1929)} \\ 
\hline
\multicolumn{6}{l}{$^{a}=$ tied between all branches, $^{b}=$ tied between HB and VX spectra}\\
\end{tabular}

\end{center}

\medskip
\tablecomments{Errors are reported at the 90\% confidence level. \nicer\ is fit in the $0.5-10$ keV energy band while \nustar is fit in the $3-30$ keV band.  A multiplicative constant is used on the \nicer and FPMB data, while FPMA is fixed to unity. The outer disk radius is fixed at 1000 \rg and the dimensionless spin parameter is set to $a_{*}=0$ (hence, 1 \risco = 6 \rg). The density in the {\sc mekal} model is fixed at $10^{15}$ cm$^{-3}$.
The unabsorbed $0.5-50$~keV flux, $F_{\rm unabs,\ 0.5-50\ keV}$, is given in units of $10^{-8}$ \fluxcgs.}

\end{rotatetable*}
\end{table*}
\movetabledown=60mm

\begin{table*}
\begin{rotatetable*}
\caption{Reflection Modeling from Comptonization}
\label{tab:nthrefl} 

%VX_NB_HB_crabcor_tbfeo_disknthcomprfxconv_ninu_chi_06282021_5E4_burn1E6.out
%VX_NB_HB_crabcor_tbfeo_disknthcomprfxconv_gauss_ninu_chi_12022021_5E4_burn1E6.out
%VX_NB_HB_crabcor_tbfeo_disknthcomprfxconv_mekal_ninu_chi_12022021_5E4_burn1p5E6.out

\begin{center}
\begin{tabular}{llccc|ccc|ccc}
\hline

Model & Parameter & \multicolumn{3}{c}{RFX1} & \multicolumn{3}{c}{RFX2} & \multicolumn{3}{c}{RFX3}\\
&& NB & VX & HB & NB & VX & HB & NB & VX & HB\\
\hline

{\sc crabcor} %%%%
& $C_{\rm FPMB}$ 
&$ 1.022 _{- 0.001 }^{+ 0.002 }$
&$ 1.014\pm0.002$
&$ 1.014 _{- 0.002 }^{+ 0.001 }$

&$ 1.022\pm0.001$%$ 1.022\pm0.001$
&$ 1.013\pm0.002$%$ 1.014 _{- 0.001 }^{+ 0.002 }$
&$ 1.015 _{- 0.003 }^{+ 0.001 }$%$ 1.014 _{- 0.002 }^{+ 0.001 }$

& $ 1.023 _{- 0.002 }^{+ 0.001 }$%$ 1.022 _{- 0.001 }^{+ 0.002 }$
& $ 1.014 _{- 0.002 }^{+ 0.001 }$%$ 1.013 _{- 0.002 }^{+ 0.002 }$
&$ 1.014 \pm 0.002$ %$ 1.014 _{- 0.002 }^{+ 0.003 }$
\\

& $C_{\rm NICER}$ 
&...
& $ 0.964 _{- 0.006 }^{+ 0.013 }$
&$ 0.991 _{- 0.007 }^{+ 0.014 }$
&...
&$ 0.969 _{- 0.013 }^{+ 0.009 }$%$ 0.97 \pm0.01$
&$ 0.998 _{- 0.015 }^{+ 0.007 }$%$ 1.00 \pm0.01$
&...
&$ 0.97 \pm0.01$%$ 0.97 \pm0.01$
&$ 1.00\pm0.01$%$ 1.00 \pm0.01$
\\

& $\Delta \Gamma_{\rm NICER}\ ^{b}$ ($10^{-2}$)
&...
& \multicolumn{2}{c|}{$ -6.1 _{- 0.3 }^{+ 0.9 }$}
&...
&\multicolumn{2}{c|}{ $ -5.6 _{- 0.9 }^{+ 0.4 }$} %\multicolumn{2}{c|}{$ -5.6 _{- 0.6 }^{+ 0.5 }$}
&...
&\multicolumn{2}{c}{$ -5.6 _{- 0.8 }^{+ 0.5 }$} %\multicolumn{2}{c}{$ -5.7 _{- 0.5 }^{+ 0.8 }$}
\\

{\sc tbfeo} %%%%
& $\mathrm{N}_{\mathrm{H}}\ ^{a}$ ($10^{21}$ cm$^{-2}$) 
& ---
&$ 2.63 _{- 0.13 }^{+ 0.05 }$
&---
&---
&$ 2.59 \pm 0.09 $ %$ 2.66 _{- 0.06 }^{+ 0.07 }$
&---
&---
&$ 2.51 _{- 0.02 }^{+ 0.13 }$%$ 2.62 \pm0.08$
&---

\\
& $A_{\rm O}\ ^{a}$
&---
&$ 1.46 _{- 0.06}^{+ 0.09}$
&---
&---
& $ 1.46 _{- 0.10 }^{+ 0.07 }$ %$ 1.43 _{- 0.07 }^{+ 0.05 }$
&---
&---
&$ 1.49 _{- 0.09 }^{+ 0.05 }$%$ 1.43 _{- 0.05 }^{+ 0.10 }$
&---

\\

{\sc diskbb} %%%%
& $kT_{\rm in}$ (keV) 
&$ 1.78 \pm0.01$
&$ 1.718 _{- 0.01 }^{+ 0.03 }$
&$ 1.69 _{- 0.01 }^{+ 0.03 }$

&$ 1.76 _{- 0.03 }^{+ 0.05 }$%$ 1.79 _{- 0.03 }^{+ 0.02 }$
&$ 1.74 _{- 0.01 }^{+ 0.02}$ %$ 1.74 _{- 0.01 }^{+ 0.02 }$
&$ 1.71 _{- 0.04 }^{+ 0.01}$ %$ 1.70 _{- 0.03 }^{+ 0.02 }$

&$ 1.78 _{- 0.04 }^{+ 0.01 }$%$ 1.75 _{- 0.01 }^{+ 0.04 }$
&$ 1.74 _{- 0.03 }^{+ 0.01 }$%$ 1.73 _{- 0.01 }^{+ 0.03 }$
&$ 1.70 \pm0.03$%$ 1.68 _{- 0.02 }^{+ 0.04 }$
\\

& norm$_{\rm disk}$ 
&$ 85 \pm3$
&$ 78 _{- 5 }^{+ 1}$
&$ 51 _{- 3 }^{+ 2 }$

& $ 86 _{- 7 }^{+ 5 }$ %$ 84 _{- 6 }^{+ 4 }$
& $ 76 _{- 4 }^{+ 3}$%$ 76 _{- 4 }^{+ 3 }$
&$ 49 _{- 1 }^{+ 5 }$%$ 50 _{- 1 }^{+ 3 }$

&$ 83 _{- 4 }^{+ 6 }$%$ 90 _{- 9 }^{+ 2 }$
&$ 77 _{- 2 }^{+ 5}$%$ 78 _{- 4 }^{+ 3 }$
&$ 51 \pm3$%$ 53 _{- 4 }^{+ 3 }$
\\

{\sc nthcomp} %%%%
& $\Gamma$
&$ 2.02 \pm0.03$
&$ 1.66 _{- 0.02 }^{+ 0.05 }$
& $ 1.73 _{- 0.01 }^{+ 0.02 }$

&$ 2.00 _{- 0.05 }^{+ 0.09 }$%$ 2.04 \pm0.05$
& $ 1.67 _{- 0.02 }^{+ 0.04 }$%$ 1.67 _{- 0.02 }^{+ 0.03 }$
& $ 1.74 _{- 0.03 }^{+ 0.01 }$%$ 1.73 _{- 0.01 }^{+ 0.02 }$

&$ 2.04 _{- 0.08 }^{+ 0.02 }$%$ 1.96 _{- 0.02 }^{+ 0.09 }$
&$ 1.65 \pm0.03$%$ 1.65 _{- 0.03 }^{+ 0.04 }$
&$ 1.72 _{- 0.02 }^{+ 0.03 }$%$ 1.71_{- 0.02 }^{+ 0.03 }$
\\

& $kT_{e}$ (keV)
&$ 2.85 _{- 0.03 }^{+ 0.07 }$
&$ 2.80_{- 0.01 }^{+ 0.08 }$
&$ 2.88 _{- 0.02 }^{+ 0.05 }$

&$ 2.75 _{- 0.05 }^{+ 0.24 }$%$ 2.91 _{- 0.14 }^{+ 0.08 }$
&$ 2.87 \pm0.04$%$ 2.85 \pm0.04$
&$ 2.93 _{- 0.05 }^{+ 0.04 }$%$ 2.90 _{- 0.05 }^{+ 0.03 }$

&$ 2.83 _{- 0.09 }^{+ 0.12}$%$ 2.73 _{- 0.04 }^{+ 0.14 }$
&$ 2.83 _{- 0.08 }^{+ 0.05 }$%$ 2.82 _{- 0.04 }^{+ 0.05 }$
&$ 2.90 _{- 0.03 }^{+ 0.05 }$%$ 2.87 _{- 0.03 }^{+ 0.06 }$
\\

& $kT_{bb}$ ($10^{-1}$~keV) 
&$ 1.0 \pm0.2$
&$ 0.8 _{- 0.1 }^{+ 0.2 }$
&$ 0.95 _{- 0.03 }^{+ 0.22 }$

&$ 1.1 _{- 0.3 }^{+ 0.2 }$%$ 1.0 _{- 0.1 }^{+ 0.3 }$
&$0.85^{\dagger}$%$ 0.06 _{- 0.01 }^{+ 0.02 }$
&$ 1.0 \pm0.2$%$ 0.9 _{- 0.2 }^{+ 0.1 }$

&$ 1.39 _{- 0.03 }^{+ 0.58 }$%$ 1.68 _{- 0.31 }^{+ 0.09 }$
&$0.85^{\dagger}$%$ 0.02 \pm0.01$
&$ 1.11 _{- 0.22 }^{+ 0.07 }$%$ 0.89 _{- 0.07 }^{+ 0.15 }$
\\

& norm$_{\rm nth}$ 
&$ 2.8 \pm0.2$
&$ 1.0 \pm0.1$
&$ 1.30 _{- 0.08 }^{+ 0.06 }$

&$ 2.9 \pm0.3$%$ 2.9 _{- 0.2 }^{+ 0.4 }$
&$ 0.96 _{- 0.06 }^{+ 0.13 }$%$ 0.99 _{- 0.07 }^{+ 0.09 }$
&$ 1.29 _{- 0.11 }^{+ 0.05 }$%$ 1.27 \pm0.05$

&$ 2.85 _{- 0.47 }^{+ 0.09 }$%$ 2.43 _{- 0.05 }^{+ 0.51 }$
&$ 0.91 _{- 0.08 }^{+ 0.10}$%$ 0.92 _{- 0.07 }^{+ 0.09 }$
&$ 1.21 _{- 0.04 }^{+ 0.10 }$%$ 1.21 _{- 0.07 }^{+ 0.08 }$
\\

%rdblur*rfxconv
{\sc rdblur} %%%%
& $|{\rm -Betor10}|$
&$ 1.5 _{- 0.5 }^{+ 0.1 }$
&$ 1.5 _{- 0.5 }^{+ 0.1 }$
&$ 1.6 _{- 0.3 }^{+ 0.1 }$

&$ 2.3 _{- 0.8 }^{+ 0.2 }$%$ 1.4 _{- 0.2 }^{+ 0.8 }$
&$ 1.3 _{- 0.3 }^{+ 0.1 }$%$ 1.1 _{- 0.1 }^{+ 0.4 }$
&$ 1.2 _{- 0.2 }^{+ 0.3 }$%$ 1.6 _{- 0.3 }^{+ 0.2 }$

&$ 1.9 _{- 0.6 }^{+ 0.2 }$%$ 2.0 _{- 0.2 }^{+ 0.1 }$
&$ 1.4 _{- 0.2 }^{+ 0.4}$%$ 1.3 _{- 0.3 }^{+ 0.2 }$
&$ 1.4 _{- 0.2 }^{+ 0.3 }$%$ 1.5 \pm0.2$
\\
& \rin\ (\rg)
&$ 6.7 \pm0.7$
&$ 6.4 _{- 0.4 }^{+ 0.6 }$
&$ 6.3 _{- 0.3 }^{+ 1.3 }$

&$ 6.6 _{- 0.6 }^{+ 1.3}$%$ 6.5 \pm0.5$
&$ 6.6 _{- 0.6 }^{+ 0.8 }$ %$ 6.06 _{- 0.06 }^{+ 1.66 }$
&$ 6.5 _{- 0.5 }^{+ 0.7 }$%$ 6.7 _{- 0.7 }^{+ 0.2 }$

&$ 6.3 _{- 0.3}^{+ 1.1 }$%$ 6.4 _{- 0.4 }^{+ 0.5 }$
&$ 6.3 _{- 0.3 }^{+ 1.1 }$%$ 6.4 _{- 0.4 }^{+ 0.5 }$
&$ 7.0 _{- 1.0 }^{+ 1.5 }$%$ 6.4 _{- 0.4 }^{+ 0.5 }$
\\
& $i\ ^{a}$ ($^{\circ}$) % 50--70 ::: 60+/-10
&---
&$ 63 _{- 3 }^{+ 4 }$
&---
&---
&$ 61 _{- 11 }^{+ 2 }$%$ 54 _{- 3}^{+ 11 }$
&---
&---
&$ 64 _{- 7 }^{+ 6 }$%$ 67 _{- 8 }^{+ 4 }$
&---
\\
{\sc rfxconv} %%%% 
& $rel_{\mathrm{refl}}$
&$ 1.0 _{- 0.2 }^{+ 0.1 }$
&$ 0.27 \pm0.05$
&$ 0.14 _{- 0.02 }^{+ 0.04 }$

&$ 1.1 _{- 0.3 }^{+ 0.1}$%$ 0.9 _{- 0.1 }^{+ 0.4 }$
&$ 0.25 _{- 0.05 }^{+ 0.02 }$%$ 0.25 _{- 0.05 }^{+ 0.06 }$
&$ 0.12 _{- 0.02 }^{+ 0.04 }$%$ 0.13 \pm0.02$

&$ 1.2 _{- 0.3 }^{+ 0.1 }$%$ 1.2 _{- 0.2 }^{+ 0.1 }$
&$ 0.26 _{- 0.06 }^{+ 0.07 }$%$ 0.23 _{- 0.03 }^{+ 0.01 }$
&$ 0.14 _{- 0.03 }^{+ 0.01 }$%$ 0.13 _{- 0.01 }^{+ 0.02 }$
\\
&$A_{\rm Fe}\ ^{a}$ 
&---
&$ 2.0 _{- 0.4 }^{+ 0.2 }$
&---
&---
&$ 1.8 _{- 0.5 }^{+ 0.2 }$ %$ 1.67 _{- 0.08 }^{+ 0.39 }$
&---
&---
&$ 1.7 _{- 0.2 }^{+ 0.5 }$%$ 2.1 _{- 0.6 }^{+ 0.2 }$
&---
\\
& $\log(\xi)$ 
&$ 2.15 _{- 0.05 }^{+ 0.17 }$
&$ 2.67 _{- 0.17 }^{+ 0.04 }$
&$ 2.83 _{- 0.12 }^{+ 0.07 }$

&$ 1.8 _{- 0.1 }^{+ 0.5 }$%$ 2.2 _{- 0.2 }^{+ 0.1 }$
&$ 2.6 \pm0.1$%$ 2.5 _{- 0.1 }^{+ 0.2 }$
&$ 2.9 \pm0.1$%$ 2.8 \pm0.1$

&$ 2.11 _{- 0.07 }^{+ 0.15 }$%$ 2.03 _{- 0.01 }^{+ 0.13 }$
&$ 2.60 _{- 0.08 }^{+ 0.13 }$%$ 2.65 _{- 0.17 }^{+ 0.06 }$
&$ 2.81 _{- 0.05 }^{+ 0.11 }$%$ 2.84 \pm0.08$
\\
{\sc Gaussian} %%%%
& E (keV)
&...
&...
&...
& ...
&$ 1.02 _{- 0.05 }^{+ 0.08 }$%$ 1.06 _{- 0.09 }^{+ 0.05 }$
&$ 1.03 _{- 0.06 }^{+ 0.03 }$%$ 0.9 _{- 0.01 }^{+ 0.11 }$
&...
&...
&...
\\
& $\sigma$ ($10^{-2}$~keV)
&...
&...
&...
& ...
&$ 6 \pm 2$%$ 8 _{- 3 }^{+ 2 }$
&$ 5 \pm1 $%$ 6.4 _{- 3.5}^{+ 0.7 }$
&...
&...
&...
\\
& norm$_{\rm gauss}$ ($10^{-2}$)
&...
&...
&...
&...
&$ 1.8 _{- 0.4 }^{+ 0.3 }$%$ 1.8 _{- 0.5 }^{+ 0.6 }$
&$ 1.3 _{- 0.2 }^{+ 0.3 }$%$ 1.4 _{- 0.5 }^{+ 0.2 }$
&...
&...
&...
\\
& EW (eV)
&...
&...
&...
& ...
&$7_{-1}^{+2}$
&$6\pm1$
&...
&...
&...
\\
{\sc mekal} %%%%
& $kT$ (keV)
&...
&...
&...
&...
&...
&...
&$ 1.9 _{- 0.6 }^{+ 0.2 }$%$ 1.6 _{- 0.2 }^{+ 0.3 }$
&$ 1.1 \pm0.3$%$ 1.1 \pm0.1$
&$ 1.5 _{- 0.4 }^{+ 0.3 }$%$ 1.1 _{- 0.1 }^{+ 0.4 }$
\\
& A$_{\rm Fe}$
&...
&...
&...
&...
&...
&...
&---
&$ 1.9 _{- 0.5 }^{+ 0.6 }$%$ 2.3 _{- 0.4 }^{+ 0.1 }$
&---
\\
& norm$_{\rm mekal}$  ($10^{-2}$)
&...
&...
&...
&...
&...
&...
&$ 0.9 _{- 0.3 }^{+ 0.6 }$%$ 1.0 _{- 0.1 }^{+ 0.2 }$
&$ 4.3 _{- 1.2 }^{+ 0.9 }$% $4.3 _{- 0.5 }^{+ 1.5 }$
&$ 4.1 _{- 1.2 }^{+ 1.0 }$%$ 3.7 _{- 0.5 }^{+ 0.6 }$
\\
&$F_{\rm unabs,\ 0.5-50\ keV}$
&$3.4\pm0.3$
&$2.3_{-0.3}^{+0.2}$
&$1.9_{-0.2}^{+0.1}$

&$3.4\pm0.4$%$3.4_{-0.3}^{+0.5}$
&$2.3\pm0.5$%$2.4_{-0.7}^{+0.8}$
&$1.9_{-0.3}^{+0.5}$%$1.9_{-0.7}^{+0.3}$

&$3.3_{-1.2}^{+2.2}$%$ 3.3_{-0.5}^{+1.0}$
&$2.3_{-0.7}^{+0.6}$%$ 2.3_{-0.3}^{+0.9}$
&$1.9_{-0.6}^{+0.5}$%$ 1.9_{-0.3}^{+0.4}$
\\

\hline
&$\chi^{2}$ (dof) & \multicolumn{3}{c}{2024.9 (1938)} & \multicolumn{3}{c}{2009.7 (1933)} & \multicolumn{3}{c}{ 2006.4 (1932)} \\ 
\hline
\multicolumn{6}{l}{$^{a}=$ tied between all branches, $^{b}=$ tied between HB and VX spectra, $^{\dagger}=$ fixed}\\
\end{tabular}
\end{center}

\medskip
\tablecomments{Errors are reported at the 90\% confidence level. \nicer\ is fit in the $0.5-10$ keV energy band while \nustar is fit in the $3-30$ keV band.  A multiplicative constant is used on the \nicer and FPMB data, while FPMA is fixed to unity. The outer disk radius is fixed at 1000 \rg and the dimensionless spin parameter is set to $a_{*}=0$ (hence, 1 \risco = 6 \rg = 12.4~ km). The density in the {\sc mekal} model is fixed at $10^{15}$ cm$^{-3}$.
The unabsorbed $0.5-50$~keV flux, $F_{\rm unabs,\ 0.5-50\ keV}$, is given in units of $10^{-8}$ \fluxcgs. The inclination parameter is tied between the {\sc rdblur} and {\sc rfxconv} convolutions. }

\end{rotatetable*}
\end{table*}

Conversely, when starting from C2 we use the reflection convolution model {\sc rfxconv} \citep{done06} with the relativistic blurring kernal {\sc rdblur} \citep{fabian89} to emulate reprocessed emission from a Comptonization blackbody with general and special relativistic effects around a non-spinning compact object (i.e., $a=0$). 
We choose {\sc rdblur} so that this overall model and RNS1 are completely independent of each other.
{\sc rfxconv} generates an angle-dependent reflection spectrum from the {\sc nthcomp} input spectrum it is convolved with by combining the Compton reflected emission from {\sc pexriv} \citep{mz95} above 14~keV (using the average $12-14$~keV power-law index) with reflection emission from an ionized disk interpolated from {\sc reflionx} \citep{ross05} below 14~keV (using the average $2-10$~keV power-law index). The reflected emission interpolated from {\sc reflionx} and {\sc pexriv} are scaled to match at 14~keV. The parameters of this convolution model are the relative reflection normalization ($rel_{\mathrm{refl}}$), the Fe abundance ($A_{Fe}$), the inclination angle ($\cos{(i)}$), redshift ($z$), and ionization parameter ($\log(\xi)$).  The parameters of the relativistic blurring kernal {\sc rdblur} are the emissivity index (Betor10: $R^{\mathrm{Betor10}}$), inner disk radius in \rg, outer disk radius, and inclination ($i$). The outer disk radius is fixed at 1000 \rg\ to be consistent with \relxillns and the inclination parameters are tied between {\sc rdblur} and {\sc rfxconv} for consistency. 
We report the parameter values for this fit in Table~\ref{tab:nthrefl} under ``RFX1".

\movetableright=-15mm
\begin{table*}[t!]
\caption{Radial Estimates from Spectral Modeling}
\label{tab:rin} 

\begin{center}
\begin{tabular}{lccccccc}
\hline
Model & Branch & $R_{\rm in,\ diskbb}$ & $R_{\rm bbody,\ spherical}$  & $R_{\rm bbody,\ banded}$  & $R_{\rm in,\ reflection}$ & $R_{\rm BL,\ max}$\\
\hline
C1 & NB & $46\pm16$ & $6\pm2$& $19\pm6$ & ... & $68\pm23$\\
& VX & $39\pm13$ & $6\pm2$ & $19\pm3$ & ... & $26\pm9$ \\
& HB & $33\pm11$ &$6\pm2$ & $20\pm7$ & ... &$19\pm7$ \\

C2 & NB &  $39\pm15$ &... & ... & ... & $42 \pm14$\\
& VX & $31\pm12$ &...& ... & ... & $32\pm11$\\ 
& HB & $24\pm9$ & ...& ... & ... & $27\pm11$\\

RNS1 & NB & $45\pm15$ & ... & ...& $15.6\pm2.0$ & $50\pm18$\\
& VX &$41\pm14$ &... &... & $16.3\pm2.2$& $25\pm12$\\
& HB &$37\pm13$ & ...& ...& $15.9\pm2.1$& $19\pm7$\\
RNS2 & NB & $46\pm15$ &... & ...&$16.3\pm2.2$ & $48\pm18$\\
& VX &$41\pm14$ &... & ...& $16.3\pm2.3$ & $25\pm9$\\
& HB &$39_{-14}^{+13}$ & ...&... & $16.3\pm2.2$ & $19\pm7$\\
RNS3 & NB &$45\pm15$ &... & ...& $15.7\pm2.0$& $50\pm21$\\
& VX &$40\pm14$ & ...&... &$16.3\pm2.2$ &$25\pm9$ \\
& HB &$38\pm13$ &... &... & $16.0\pm2.1$ & $18\pm7$\\

RFX1 & NB & $34\pm13$ &... &... & $16.9\pm2.7$& $48\pm16$\\
& VX & $33\pm12$  &... &... &$16.1^{+2.5}_{-2.2}$ & $32\pm11$\\
& HB & $27\pm10$ &... &... &$15.9_{-2.1}^{+3.8}$ & $27\pm9$\\

RFX2  & NB & $35\pm13$  &... &... & $16.6_{-2.5}^{+3.9}$ & $48\pm17$\\
& VX &$33\pm12$ &... &... &$16.6_{-2.5}^{+2.9}$ & $32\pm13$\\
& HB & $26\pm10$ &... &... & $16.4_{-2.4}^{+2.7}$ & $27\pm11$\\

RFX3 & NB & $34\pm13$ &... &... & $15.9_{-2.1}^{+3.4}$ & $46_{-23}^{+34}$ \\
& VX & $33\pm12$ &... &... & $15.9_{-2.1}^{+3.4}$  & $32\pm14$ \\
& HB & $27\pm10$ &... &... & $17.6_{-3.3}^{+4.4}$  & $27\pm12$ \\
\hline

\end{tabular}
\end{center}

\medskip
Note.--- All values are given in units of km. The inner disk radius from reflection modeling was converted into km assuming a NS mass of $M_{\rm NS}=1.71\pm0.21$~\msun (and $a_{*}=0$ in the case of values taken from Table 3). Estimates encompass the entire reported distance range to the source ($9.15\pm3.05$~kpc). A color correction factor of $f_{cor}=1.7$ \citep{shimura95}  was used when converting the normalization of {\sc diskbb} and {\sc bbody} components into their emitting radius. $R_{\rm BL,\ max}$ is the radial extent of the boundary layer from the surface of the NS using equation (25) from \citet{PS01}. For C1 and fits labeled RNS, the inclination range from fitting {\sc relxillNS} was used ($i=67^{\circ}\pm4^{\circ}$). For C2 and fits labeled RFX, the inclination range from fitting {\sc rdblur*rfxconv} was used ($i=60^{\circ}\pm10^{\circ}$).

\end{table*}

The 1~keV feature is still present in the spectrum regardless of the reflection model utilized (see Figure \ref{fig:ratios} panels (c) and (d)). We proceed to model the 1~keV feature in the \nicer data with a Gaussian emission line ({\sc gauss}) to determine the line centroid energy and equivalent width. These fits are reported as ``RNS2" and ``RFX2" in Table~\ref{tab:bbrefl} and \ref{tab:nthrefl} while the ratio of the overall model to the data are shown in Figure \ref{fig:ratios}(e) and (f), respectively. The addition of the Gaussian line improved the fit by $8.2\sigma$ when using \relxillns, but provides a marginal  2.5$\sigma$ improvement from RFX1 to RFX2.  We note that the seed photon temperature of the Comptonization model tends to an unphysically low value ($kT_{bb}\leq8$~eV) in the VX branch with the addition of the Gaussian line, therefore we fix this parameter to the median value from RFX1. This was not an issue in the HB due to the \nicer\ spectrum containing $4.8\times10^{6}$ more counts than in the VX. All parameter values are consistent within the 90\% confidence level when $kT_{bb}$ is left free or fixed, hence this parameter does not have a notable impact on the results.  
Although the feature is present regardless of continuum and reflection modeling, the inferred equivalent width is smaller in RFX2 which may be due to the low-energy turnover in the Comptonization model. 
Additionally, the strength of the emission line does appear to change with flux as reported by \citet{vrtilek86}. Indeed, the energy of this line is too low to originate from the relativistic reflection component like the Fe~L emission in Serpens~X-1 \citep{ludlam18}. It could be a blend of Fe, Ni, and O transition lines that originate from collisionally ionized material far from the inner region of the accretion disk as proposed by \citet{vrtilek86}. 

We proceed to replace the Gaussian emission line component for a collisionally ionized plasma model {\sc mekal} \citep{mewe85,mewe86,liedhal95} to determine how this interpretation impacts the inferred inner disk radius. The density of the material is fixed at $10^{15}$~cm$^{-3}$  \citep{schulz09} and the abundance of the plasma is tied between the spectra. The temperature and normalization is free to vary. This fit is referred to as ``RNS3" and ``RFX3" in Tables \ref{tab:bbrefl} and \ref{tab:nthrefl}. 
Again, the seed photon temperature of the Comptonization component tended to an unphysical value of $kT_{bb}\leq3$~eV in the RFX3 VX branch and was fixed at the median value from RFX1. However, the results with this parameter fixed agrees within the 90\% confidence level when $kT_{bb}$ was free to vary.
The ratio of the model to the data are shown in Figure \ref{fig:ratios}(g) and (h). The addition of a {\sc mekal} component represents a $7.6\sigma$ and 2.7$\sigma$ improvement in comparison to RNS1 and RFX1, respectively. Figure \ref{fig:ratios}(i) and (j) show the unfolded spectrum with the model components for RNS3 and RFX3, respectively. The {\sc mekal} model predicts a narrow emission line in the Fe~K band as well, but this is orders of magnitude below the broadened emission line from reflection. The normalization of the {\sc mekal} component for the NB in RFX3 is lower in comparison to the VX and HB, but this could be due to the lack of \nicer data to anchor the component through modeling of the 1~keV feature. The exact nature of the 1~keV component is beyond the scope of this paper, but regardless of how the feature is modeled the inner disk radius still remains close to the NS.

\begin{figure*}[t!]
%\vspace{-7pt}
\begin{centering}
\includegraphics[width=18.2cm, trim=20 0 0 0, clip]{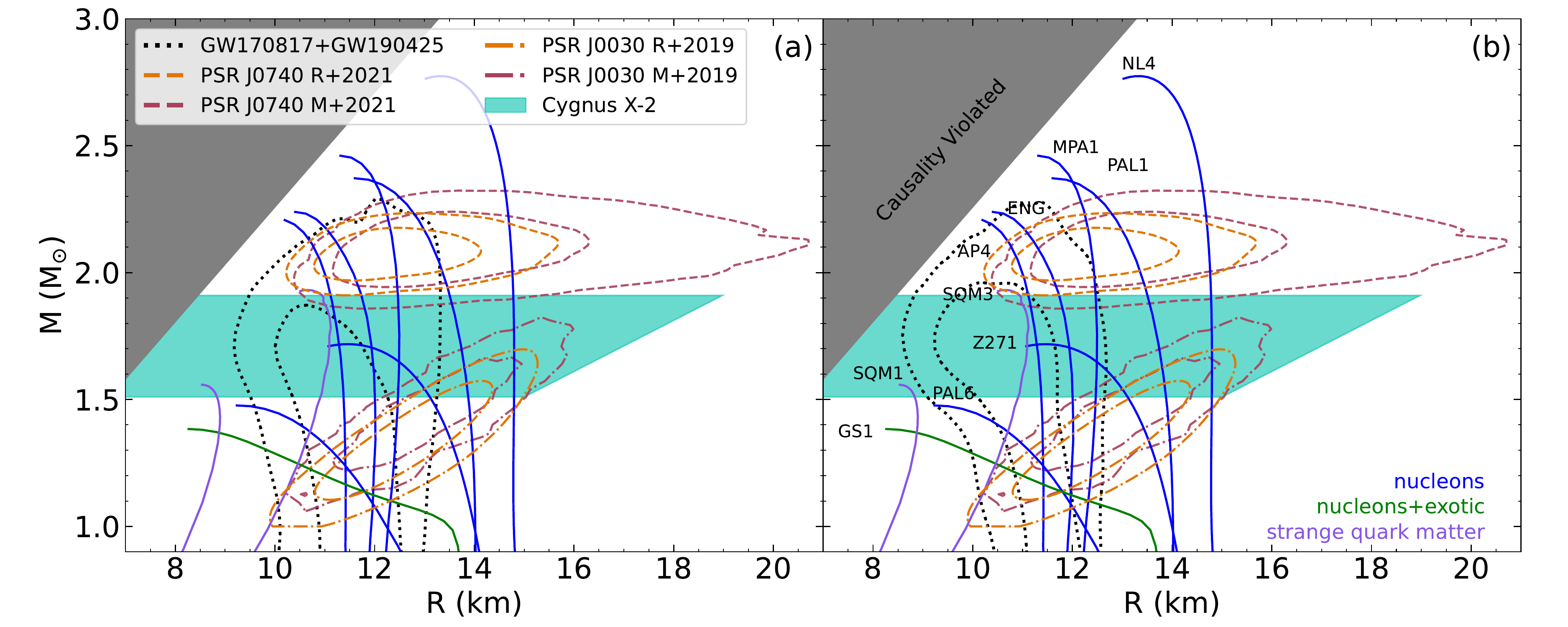}
\caption{Mass and radius constraints from NS gravitational wave events and \nicer pulsar light curve modeling in comparison to reflection modeling of \source. Panels differ by the $M$--$R$ constraints from the combined GW170817 \citep{abbott19} and GW190425 \citep{gw19} signatures using a (a) piece-wise polytropic model and (b) speed-of-sound model through a Bayesian framework as reported in \citet{raaijmakers21}. These are denoted by the black dotted line. The solid cyan region indicates the conservative radiusconstraints for \source based on the reflection modeling reported in Table~\ref{tab:bbrefl}. %The lighter hatched region indicates values less than 1~\risco, which is the hard-coded lower bound of the reflection model. 
The solid grey region indicates where causality is violated (i.e., sound speed within the NS exceeds the speed of light). Select EoS from \citet{LP01} are shown to demonstrate the behavior on the $M$--$R$ plane for a given internal composition. Pulsar light curve modeling of \nicer data for PSR J0740+6620 are indicated by dashed lines from \citet{riley21} (R+2021: orange) and \citet{miller21} (M+2021: maroon). 
The dot-dashed lines indicate the results for light curve modeling of PSR~J0030+0452 reported in \citet{riley19} (R+2019: orange) and \citet{miller19} (M+2019: maroon). Confidence contours correspond to the 68\% and 95\% credible regions.
}
\label{fig:MR}
\end{centering}
\end{figure*} 

\section{Discussion}
We present an analysis of the reflection spectrum in \source using three \nustar and two simultaneous \nicer observations. The source traced out the flaring to horizontal branch within these observations. The data were divided into the respective branches and spectra with $\geq10^{6}$ cumulative counts were modeled according to different continuum conventions. This resulted in a spectrum of the source in the normal branch (NB) from the first \nustar observation, the vertex (VX) between the normal and horizontal branch (HB) during the second \nustar observation, and the HB in the third \nustar observation. Simultaneous \nicer spectra were extracted for the VX and HB allowing for spectral modeling from $0.5-30$~keV. The reflection spectrum was modeled with \relxillns and {\sc rfxconv} depending on the illuminating continuum component. Regardless of which reflection model was utilized, the inner disk radius remained close to \risco (1~\risco = 6~\rg for $a=0$). 

The inferred inclination from reflection modeling (\relxillns: $i=67^{\circ}\pm4^{\circ}$, {\sc rdblur*rfxconv}: $i=60^{\circ}\pm10^{\circ}$) is consistent with the optical results ($i=62.5^{\circ}\pm4^{\circ}$: \citealt{orosz99}), but conflicts with the previously reported low inclinations from reflection modeling in \citet{cackett10} and \citet{mondal18}. Again, one possible explanation could be the material further out partially obscuring the blue-wing emission of the Fe~K line \citep{taylor18} at the time of the {\it Suzaku} observation reported in \citet{cackett10}. This effect has been invoked to explain conflicting inclination measurements between reflection modeling and dynamical estimates in the black hole X-ray binary XTE J1550-564 \citep{connors19}. In the case of the \citet{mondal18} study that utilized the same \nustar observation from 2015, the differences between the results reported therein and here could be due to differences in how the data were reduced (e.g., bright source flag expression in `nupipeline') and handled (e.g., our self-consistent reflection modeling and choice to tie various parameters across observations). The inclination varies more when using \rfxconv than from \relxillns, but this is likely due to difference in the relativistic convolution routines within each model as discussed in \citet{ludlam20}. Further differences between the models may be due to the hard coded disk density in \rfxconv of $10^{15}$ cm$^{-3}$, while the \relxillns model has a variable disk density component that allows for conditions closer to the physical density expected in accretion disk of LMXBs. The hard-coded lower disk density in {\sc rfxconv} is likely responsible for the higher inferred iron abundance, $A_{\rm Fe}$, in comparison to the results from \relxillns which are closer to solar abundances.

The emissivity indices are lower than the $q=3$ profile for Euclidean geometry, but are close to the expected shallower illumination profile from an extended disk corona around a slowly spinning compact object \citep{kinch16, kinch19}.
The ionization parameter is consistent with the value reported in \citet{cackett10}, but higher than those found in \citet{mondal18}. The \citet{mondal18} study also found subsolar Fe abundances, which may explain the lower inferred ionization since positive correlations between $\log(\xi)$ and $A_{\rm Fe}$ have been observed previously when modeling reflection in X-ray binaries (e.g., \citealt{connors19}). Our choice to tie the $A_{\rm Fe}$ between observations reduces the degeneracy while allowing the $\log(\xi)$ to be a free parameter. 

\subsection{Radius Constraints}
Apart from the inner disk radius returned by the reflection model components, we can calculate the inferred emission radius from the normalizations of the thermal disk component ({\sc diskbb}) and single-temperature blackbody ({\sc bbody}). The emitting blackbody radius is given assuming both a spherical and a narrow banded emission region on the surface of the NS with a vertical height that is 10\% of the radial extent \citep{PS01,ludlam21}. Additionally, we can calculate the maximum radial extent of a boundary layer extending from the surface of the NS based on the mass accretion rate from Equation (25) in \citet{PS01} for comparison. It is important to note that the maximum radial extent using this equation does not account for spreading of the boundary layer in the vertical direction or effects from rotation of the NS. We report these values in Table \ref{tab:rin} in units of km for each spectral model reported in Tables \ref{tab:continuum} -- \ref{tab:nthrefl} and branch. The implausibly small inferred emitting radius of the blackbody component when assuming spherical emission supports the presence of a narrow banded emission region \citep{inogamov99}. The inner disk radius inferred from the disk component is larger than that inferred from the reflection modeling. The choice of spectral hardening factor does impact the inferred radius (e.g., \citealt{kubota01}). 
However, the inferred {\sc diskbb} radius is also known to be up to a factor of $\sim2.2$ smaller when accounting for zero-torque inner boundary condition expected for thin disk accretion \citep{zimmerman05}.  This brings the inner disk radius values from the {\sc diskbb} component within the uncertainty of the reflection model. The relative agreement between the inferred radius measurements of each of these components is an encouraging cross-check on the validity of the overall spectral modeling results.

\subsection{$M$--$R$ Plane}
Given that the inner accretion disk is close to the \risco in all cases, we explore the constraints that this translates to on the $M$--$R$ plane for NSs, and hence the EoS of ultradense, cold matter. The radius of the ISCO around a compact object in units of gravitational radii is dependent upon the dimensionless spin parameter, $a = cJ/GM^{2}$, of the compact object \citep{bardeen72}. The spin therefore enables a translation from ISCO to gravitational radii. This can then be converted into kilometers given the NS mass estimate of $1.71\pm0.21$ \msun for \source \citep{casares10}. 
Figure \ref{fig:MR} plots the range that the more conservative constraints from \relxillns of \rin~$\leq 1.15$ \risco for $a=0$ corresponds to on the $M$--$R$ plane used to characterize the EoS. Note that a higher spin value corresponds to a smaller gravitational radius for \risco, therefore we are presenting the most conservative upper limit when using $a=0$. 
This is then compared to $M$--$R$ estimates from gravitational wave signatures of binary NS and pulsar light-curve modeling.
The combined gravitational wave constraints from the double NS mergers GW170817 \citep{abbott19} and GW190425 \citep{gw19} are determined through a Bayesian framework as reported in \citet{raaijmakers21} for both a piece-wise polytropic model (Fig \ref{fig:MR}a) and speed-of-sound model (Fig \ref{fig:MR}b). Additionally, the \nicer pulsar light-curve modeling results for PSR J0030+0425 \citep{riley19, miller19} and PSR J0740+6620 \citep{riley21, miller21} are shown. 
The allowed region on the $M$--$R$ plane for \source could be further narrowed down in the future by obtaining improved mass constraints with better optical light curve data and ellipsoidal modeling.
Disk reflection is able to provide an upper limit on the NS radii but it is not able to rule out plausible EoSs on its own.
Each of these methods have their own systematic uncertainties (see discussions in \citealt{ludlam17a, riley19, miller19, raaijmakers21}, and references therein), but they can provide independent checks of constraints from the others on the $M$--$R$ plane. \\ 

\section{Conclusion}
We perform a spectral analysis of \nustar and \nicer observations of Cygnus~X-2 while the source was in the normal branch, vertex, and horizontal branch constrain the inner disk radius via reflection modeling. A broad Fe line component was detected in all states, as well as a 1~keV emission line where \nicer data was available. The reflection spectrum was modeled in two different ways assuming: (1) an illuminating blackbody component and (2) a Comptonization thermal component. The low-energy emission line was not able to be modeled by the reflection component suggesting that it originated further out in the accretion disk. Regardless of the reflection model utilized or how the 1~keV feature was accounted for, the inner disk radius remained close to the NS. We utilized these measurements to place an upper limit on the radius of the NS. When taken in comparison to state-of-the-art methods, disk reflection can provide an independent check of constraints on the $M$--$R$ plane. \\

{\it Acknowledgements:} Support for this work was provided by NASA through the NASA Hubble Fellowship grant \#HST-HF2-51440.001 awarded by the Space Telescope Science Institute, which is operated by the Association of Universities for Research in Astronomy, Incorporated, under NASA contract NAS5-26555. This research has made use of the NuSTAR Data Analysis Software (NuSTARDAS) jointly developed by the ASI Science Data Center (ASDC, Italy) and the California Institute of Technology (Caltech, USA).


\begin{thebibliography}{}
\bibitem[Abbott et al.(2019)]{abbott19}Abbott, B. P., Abbott, R., Abbott, T. D., et al.\ 2019, Physical Review X, 9, 011001
\bibitem[Ba\l uci\'nska-Church et al.(2010)]{bc10}Ba\l uci\'nska-Church, M., Gibiec, A., Jackson, N. K., \& Church, M. J.\ 2010, A\&A, 512, A9
\bibitem[Ba\l uci\'nska-Church et al.(2011)]{bc11}Ba\l uci\'nska-Church, M., Schulz, N. S., Wilms, J., et al.\ 2011, A\&A, 530, A102
\bibitem[Bardeen et al.(1972)]{bardeen72}Bardeen, J. M., Press, W. H., \& Teukolsky, S. A.\ 1972, ApJ, 178, 347
\bibitem[Byram et al.(1966)]{byram66}Byram E. T., Chubb T. A., Friedman H., 1966, AJ, 71, 379
\bibitem[Cackett et al.(2008)]{cackett08}Cackett, E. M., Miller, J. M., Bhattacharyya, S., et al.\ 2008, ApJ, 674, 415
\bibitem[Cackett et al.(2010)]{cackett10}Cackett, E. M., Miller, J. M., Ballantyne, D. R., et al.\ 2010, ApJ, 720, 205
\bibitem[Casares et al.(1998)]{casares98}Casares J., Charles P., Kuulkers E., 1998, ApJ, 493, L39
\bibitem[Casares et al.(2010)]{casares10}Casares J., Gonz\'alez Hern\'andez J. I., Israelian G., Rebolo R., 2010, MNRAS, 401, 2517
\bibitem[Chiappetti et al.(1990)]{chiappetti90}Chiappetti, L., Treves, A., Branduardi-Raymont, G., et al.\ 1990, ApJ, 361, 596
\bibitem[Connors et al.(2019)]{connors19} Connors, R. M. T., Garc\'ia, J. A., Steiner, J. F., et al.\ 2019, ApJ, 882, 179
\bibitem[Coughenour et al.(2018)]{coughenour18} Coughenour, B.\ M., Cackett, E.\ M., Miller, J.\ M., \& Ludlam, R. M.\ 2018, \apj, 867, 64
\bibitem[Cowley et al.(1979)]{cowley79}Cowley A. P., Crampton D., Hutchings J. B., 1979, ApJ, 231, 539
\bibitem[Di Salvo et al.(2002)]{disalvo02}Di Salvo, T., Farinelli, R., Burderi, L., et al.\ 2002, A\&A, 386, 535
\bibitem[Ding et al.(2021)]{ding21} Ding, H., Deller, A. T., \& Miller-Jones, J. C. A.\ 2021, arXiv:2105.05164
\bibitem[Done \& Gierl\'{i}nski(2006)]{done06}Done, C., \& Gierl\'{i}nski, M. 2006, MNRAS, 367, 659
\bibitem[Fabian et al.(1989)]{fabian89}Fabian, A. C., Rees, M. J., Stella, L., \& White, N. E.\ 1989, MNRAS, 238, 729
\bibitem[Fabian et al.(2000)]{fabian00}Fabian, A. C., Iwasawa, K., Reynolds, C. S., \& Young, A. J. 2000, PASP, 112, 1145
\bibitem[Farinelli et al.(2009)]{farinelli09}Farinelli, R., Paizis, A., Landi, R., \& Titarchuk, L.\ 2009, A\&A, 498, 509
\bibitem[Fridriksson et al.(2015)]{fridriksson15}Fridriksson, J. K., Homan, J., \& Remillard, R. A.\ 2015, ApJ, 809, 52
\bibitem[Galloway et. al.(2008)]{galloway08}Galloway, D. K., Muno, M. P., Hartman, J. M., Psaltis, D., \& Chakrabarty, D.\ 2008, ApJS, 179, 360
\bibitem[Garc\'{i}a et al.(2014)]{garcia14}Garc\'{i}a, J., Dauser, T., Lohfink, A., et al.\ 2014, \apj, 782, 76
\bibitem[Hasinger \& van der Klis(1989)]{HK89}Hasinger, G., \& van der Klis, M.\ 1989, A\&A, 225, 79
\bibitem[HI4PI Collaboration et al.(2016)]{HI4}HI4PI Collaboration, Ben Bekhti, N., Fl\"{o}er, L., et al.\ 2016, A\&A, 594, A116
\bibitem[Inogamov \& Sunyaev(1999)]{inogamov99}Inogamov, N. A., \& Sunyaev, R. A.\ 1999, Astron. Lett., 25, 269
\bibitem[Khan \& Grindlay(1984)]{khan84} Kahn S. M., Grindlay J. E., 1984, ApJ, 281, 826
\bibitem[Kubota et al.(2001)]{kubota01}Kubota, A., Makishima, K., \& Ebisawa, K.\ 2001, ApJ, 560, L147
\bibitem[Kinch et al.(2016)]{kinch16} Kinch, B. S., Schnittman, J. D., Kallman, T. R., \& Krolik, J. H.\ 2016, ApJ, 826, 52
\bibitem[Kinch et al.(2019)]{kinch19} Kinch, B. S., Schnittman, J. D., Kallman, T. R., \& Krolik, J. H.\ 2019, ApJ, 873, 71
\bibitem[Kuulkers et al.(1996)]{kuulkers96}Kuulkers, E., van der Klis, M., \& Vaughan, B. A.\ 1996, A\&A, 311, 197
\bibitem[Kuulkers et al.(1997)]{kuulkers97}Kuulkers, E., Parmar, A. N., Owens, A., Oosterbroek, T., \& Lammers, U.\ 1997, A\&A, 323, L29
\bibitem[Lattimer \& Prakash(2001)]{LP01}Lattimer, J. M., \& Prakash, M. 2001, ApJ, 550, 426
\bibitem[Liedhal et al.(1995)]{liedhal95} Liedahl, D. A., Osterheld, A. L., \& Goldstein, W H.\ 1995, ApJ, 438, 115
\bibitem[Lin et al.(2007)]{lin07}Lin, D., Remillard, R. A., \& Homan, J.\ 2007, \apj, 667, 1073
\bibitem[Lin et al.(2007)]{lin07}Lin, D., Remillard, R. A., \& Homan, J.\ 2007, \apj, 667, 1073
\bibitem[Ludlam et al.(2017a)]{ludlam17a}Ludlam, R.\ M., Miller, J.\ M., Bachetti, M., et al.\ 2017a, ApJ, 836, 140
\bibitem[Ludlam et al.(2018)]{ludlam18}Ludlam, R. M., Miller, J. M., Arzoumanian, Z., et al.\ 2018, ApJL, 858, L5
\bibitem[Ludlam et al.(2019)]{ludlam19}Ludlam, R.\ M., Miller, J.\ M., Barret, D., et al.\ 2019, ApJ, 873, 99
\bibitem[Ludlam et al.(2020)]{ludlam20}Ludlam, R. M., Cackett, E. M., Garc\'{i}a, J. A., et al.\ 2020, ApJ, 895, 45
\bibitem[Ludlam et al.(2021)]{ludlam21}Ludlam, R. M., Jaodand, A. D., Garc\'{i}a, J. A., et al.\ 2021, ApJ, 911, 123
\bibitem[Magdziarz \& Zdziarski(1995)]{mz95}Magdziarz, P., \& Zdziarski, A.\ 1995, MNRAS, 273, 837
\bibitem[Mewe et al.(1985)]{mewe85}Mewe, R., Gronenschild, E. H. B. M., \& van den Oord, G. H. J.\ 1985, A\&AS, 62, 197
\bibitem[Mewe et al.(1986)]{mewe86} Mewe, R., Lemen, J. R., \& van den Oord, G. H. J.\ 1986, A\&AS, 65, 511
\bibitem[Miller et al.(2011)]{miller11}Miller, J. M., Maitra, D., Cackett, E. M., Bhattacharyya, S., \& Strohmayer, T. E.\ 2011, ApJL, 731, L7
\bibitem[Miller et al.(2019)]{miller19}Miller, M. C., Lamb, F. K., Dittmann, A. J., et al.\ 2019, ApJL, 887, L24. Zenodo. https://doi.org/10.5281/zenodo.3473466
\bibitem[Miller et al.(2021)]{miller21}Miller, M. C., Lamb, F. K., Dittmann, A. J., et al.\ 2021, ApJL, 918, L28. Zenodo. https://doi.org/10.5281/zenodo.4670689
\bibitem[Mitsuda et al.(1994)]{mitsuda94}Mitsuda, K., Inoue, H., Koyama, K., et al.\ 1984, PASJ, 36, 741
\bibitem[Mondal et al.(2018)]{mondal18} Mondal, A.\ S., Dewangan, G.\ C., Pahari, M., \& Raychaudhuri, B.\ 2018, MNRAS, 474, 2064
\bibitem[Orosz \& Kuulkers(1999)]{orosz99}Orosz J. A., Kuulkers E., 1999, MNRAS, 305, 1320
\bibitem[\"Ozel \& Freire(2016)]{ozel16}\"Ozel, F., \& Freire, P.\ 2016, ARA\&A, 54, 401
\bibitem[Popham \& Sunyaev(2001)]{PS01}Popham, R., \& Sunyaev, R. 2001, ApJ, 547, 355
\bibitem[Psaradaki et al.(2020)]{psaradaki20} Psaradaki, I., Costantini, E.,  Mehdipour, M., et al.\ 2020, A\&A, 642, 208
\bibitem[Raaijmakers et al.(2021)]{raaijmakers21} Raaijmakers, G., Greif, S. K., Hebeler, K., et al.\ 2021, ApJL, 918, L29.  Zenodo. https://doi.org/10.5281/zenodo.4696232
\bibitem[Remillard et al.(2021)]{remillard21}Remillard, R. A., Loewenstein, M., Steiner, J. F., et al.\ 2021, arXiv:2105.09901
\bibitem[Riley et al.(2019)]{riley19}Riley, T. E., Watts, A. L., Bogdanov, S., et al.\ 2019, ApJL, 887, L21. Zenodo. https://doi.org/10.5281/zenodo.5506838
\bibitem[Riley et al.(2021)]{riley21}Riley, T. E., Watts, A. L., Ray, P. S., et al.\ 2021, ApJL, 918, L27. Zenodo. https://doi.org/10.5281/zenodo.4697625
\bibitem[Ross \& Fabian(2005)]{ross05}Ross, R.\ R., \& Fabian, A.\ C.\ 2005, MNRAS, 358, 211
\bibitem[Shaposhnikov et al.(2009)]{shaposhnikov09}Shaposhnikov, N., Titarchuk, L., \& Laurent, P.\ 2009, ApJ, 699, 1223
\bibitem[Shimura \& Takahara(1995)]{shimura95}Shimura, T., \& Takahara, R.\ 1995, ApJ, 445, 780
\bibitem[Schulz et al.(2009)]{schulz09}Schulz, N. S., Huenemoerder, D. P., Ji, L., et al.\ 2009, ApJ, 692, L80
\bibitem[Smale et al.(1993)]{smale93}Smale, A. P., Done, C., Mushotzky, R. F., et al.\ 1993, ApJ, 410, 796
\bibitem[Smale(1998)]{smale98}Smale A. P., 1998, ApJ, 498, L141
\bibitem[Steiner et al.(2010)]{steiner10}Steiner, J. F., McClintock, J. E., Remillard, R. A., et al.\ 2010, ApJL, 718, L117
\bibitem[Taylor \& Reynolds(2018)]{taylor18}Taylor, C., \& Reynolds, C. S.\ 2018, ApJ, 855, 120
\bibitem[The LIGO Scientific Collaboration et al.(2020)]{gw19}The LIGO Scientific Collaboration, the Virgo Collaboration, Abbott, B. P., et al.\ 2020, ApJ, 892, L3. http://dx.doi.org/10.3847/2041-8213/ab75f5
\bibitem[Verner et al.(1996)]{verner96}Verner, D. A., Ferland, G. J., Korista, K. T., \& Yakovlev, D. G.\ 1996, ApJ, 465, 487
\bibitem[Vrtilek et al.(1986)]{vrtilek86}Vrtilek, S. D., Swank, J. H., Kelley, R. L., \& Kahn, S. M.\ 1986, ApJ, 307, 69
\bibitem[Vrtilek et al.(1988)]{vrtilek88}Vrtilek, S. D., Kahn, S. M., Grindlay, J. E., Helfand, D. J., \& Seward, F. D.\ 1988, ApJ, 329, 276
\bibitem[Wijnands et al.(1997)]{wijnands97} Wijnands R., van der Klis M., Kuulkers E., Asai K., \& Hasinger G., 1997,A\&A, 323, 399
\bibitem[Wijnands \& van der Klis(2001)]{wijnands01} Wijnands, R., \& van der Klis, M.\ 2001, MNRAS, 321, 537
\bibitem[Wilms et al.(2000)]{wilms00}Wilms, J., Allen, A., \& McCray, R.\ 2000, ApJ, 542, 914
\bibitem[Zdziarski et al.(1996)]{zdziarski96}Zdziarski, A. A., Johnson, W. N., \& Magdziarz, P.\ 1996, MNRAS, 283, 193
\bibitem[Zimmerman et al.(2005)]{zimmerman05}Zimmerman, E. R., Narayan, R.,  McClintock, J. E., \& Miller, J. M.\ 2005, ApJ, 618, 832
\bibitem[Zycki et al.(1999)]{zycki99}Zycki, P. T., Done, C., \& Smith, D. A. 1999, MNRAS, 309, 561

\end{thebibliography}
\end{document}